\documentclass[]{pasj01}
\Received{2023/08/15}%{yyyy/mm/dd}
\Accepted{2023/11/07}%{yyyy/mm/dd}
\usepackage{natbib}    % reference
\usepackage{graphicx}	% Including figure files
\usepackage{comment}
\usepackage{url}
\let\leftroot\relax
\let\uproot\relax

\usepackage{amsmath}	% Advanced maths commands

\let\leftrightarrows\relax
\let\gtrsim\relax

\let\lesssim\relax

\let\varnothing\relax
\usepackage{amssymb}	% Extra maths symbols
\usepackage{threeparttable}
\usepackage[switch,mathlines]{lineno}
\newcommand{\scm}{cm$^{-2}$}	% per cm-squared
\newcommand{\ccm}{cm$^{-3}$}	% per cm-cubed
\newcommand{\HII}{H{\sc ii}}
\newcommand{\kms}{km~s$^{-1}$}
\newcommand{\co}{$^{12}$CO}
\newcommand{\cothirteen}{$^{13}$CO}
\newcommand{\coeighteen}{C$^{18}$O}
\newcommand{\ammonia}{NH$_{3}$}
\newcommand\degr{\hbox{$^\circ$}}
\newcommand\pasa{Publ. Astron. Soc. Aust}
\begin{document} 

\title{ 
%\LETTERLABEL %%% <-- uncomment for LETTER article  
%\REVIEWLABEL %%% <-- uncomment for REVIEW article  
KAgoshima Galactic Object survey with Nobeyama 45-metre telescope by Mapping in Ammonia lines (KAGONMA): Discovery of parsec-scale CO depletion in the Canis Major star-forming region}

%%% begin:list of authors
% Do NOT capitalize all letters in "textsc".
\author{Yushi \textsc{Hirata},\altaffilmark{1}$^{*}$
        Takeru \textsc{Murase},\altaffilmark{1,2}
        Junya \textsc{Nishi},\altaffilmark{1}
        Yoshito \textsc{Shimajiri},\altaffilmark{1,3,4,5}
        Toshihiro \textsc{Omodaka},\altaffilmark{1,6}
        Makoto \textsc{Nakano},\altaffilmark{7}
        Kazuyoshi \textsc{Sunada},\altaffilmark{8}
        Takumi \textsc{Ito},\altaffilmark{1,9} and
        Toshihiro \textsc{Handa}\altaffilmark{1,6}$^{*}$}

\altaffiltext{1}{Department of Physics and Astronomy, Graduate School of Science and Engineering, Kagoshima University, 1-21-35 Korimoto, Kagoshima, Kagoshima 890-0065, Japan}
\altaffiltext{2}{Faculty of Engineering, Gifu University, 1-1 Yanagido, Gifu 501-1193, Japan}
\altaffiltext{3}{National Astronomical Observatory of Japan, Osawa 2-21-1,Mitaka, Tokyo 181-8588, Japan}
\altaffiltext{4}{Laboratoire d’Astrophysique (AIM), CEA/DRF, CNRS, Universit\'{e} Paris-Saclay, Universit\'{e} Paris Diderot, Sorbonne Paris Cit\'{e}, 91191 Gif-sur-Yvette, France}
\altaffiltext{5}{Kyushu Kyoritsu University, Jiyugaoka 1-8, Yahatanishi-ku,Kitakyushu, Fukuoka, 807-8585, Japan}
\altaffiltext{6}{Amanogawa Galaxy Astronomy Research Center, Kagoshima University, 1-21-35 Korimoto, Kagoshima, Kagoshima 890-0065, Japan}
\altaffiltext{7}{Faculty of Science and Technology, Oita University, Oita 870-1192, Japan}
\altaffiltext{8}{Mizusawa VLBI observatory, NAOJ 2-12, Hoshigaoka, Mizusawa, Oshu, Iwate 023-0861, Japan}
\altaffiltext{9}{Graduate School of Science and Technology, Kumamoto University, Kumamoto, 860-8555, Japan}

\email{yushi.hirata.astro@gmail.com}
\email{k3233287@kadai.jp}
\email{handa@sci.kagoshima-u.ac.jp}
%%% end:list of authors

%% `\KeyWords{}' always has to be placed before ``\maketitle'' 
%%  List of Key Words:  https://academic.oup.com/pasj/pages/Pasj_Keywords 

\maketitle

\begin{abstract}
In observational studies of infrared dark clouds, the number of detections of CO freeze-out onto dust grains (CO depletion) at pc-scale is extremely limited, and the conditions for its occurrence are, therefore, still unknown. We report a new object where pc-scale CO depletion is expected. As a part of Kagoshima Galactic Object survey with Nobeyama 45-m telescope by Mapping in Ammonia lines (KAGONMA), we have made mapping observations of \ammonia\ inversion transition lines towards the star-forming region associated with the CMa OB1 including IRAS 07077--1026, IRAS 07081--1028, and PGCC G224.28--0.82. By comparing the spatial distributions of the \ammonia\ (1,1) and \coeighteen\ ($J$=1--0), an intensity anti-correlation was found in IRAS 07077--1026 and IRAS 07081--1028 on the $\sim$1~pc scale. Furthermore, we obtained a lower abundance of \coeighteen\ at least in IRAS 07077--1026 than in the other parts of the star-forming region. After examining high density gas dissipation, photodissociation, and CO depletion, we concluded that the intensity anti-correlation in IRAS 07077--1026 is due to CO depletion. On the other hand, in the vicinity of the centre of PGCC G224.28--0.82, the emission line intensities of both the \ammonia\ (1,1) and \coeighteen\ ($J$=1--0) were strongly detected, although the gas temperature and density were similar to IRAS 07077--1026. This indicates that there are situations where \coeighteen\ ($J$=1--0) cannot trace dense gas on the pc scale and implies that the conditional differences that \coeighteen\ ($J$=1--0) can and cannot trace dense gas are unclear.
\end{abstract}

%\pagewiselinenumbers

\section{Introduction}
Star formation is known to occur in cold, dense molecular clouds. In general, molecular clouds have a temperature of $\sim$10 K, a volume density of about 10$^2$ -- 10$^5$ \ccm\ \citep[e.g.,][]{Dame_2001,Bergin_2007}, and are mostly composed of molecular hydrogen.
The basic approach to measuring the mass of molecular clouds is to count the number of hydrogen molecules along the line-of-sight, called the H$_2$ column density, $N$(H$_2$). 
However, radiation from molecular hydrogen, which is most abundant in molecular clouds, can only be detected from gas with a temperature above 80~K \citep{Togi_2016}.
This means that it is not possible to observe H$_2$ emission from low-temperature regions directly related to the star-forming activity.
Therefore, it is often used to determine the column density of CO molecules, $N$(CO), from the emission of CO isotopologues (e.g., \co, \cothirteen, and \coeighteen), which are the most abundant molecules after H$_2$, and use the [CO]/[H$_2$] abundance ratio to measure $N$(H$_2$) indirectly.

$N$(CO) and $N$(H$_2$) are not necessarily linear relationships \citep[e.g.,][]{Frerking_1982,Pineda_2008,Ripple_2013,Wang_2019}.
In the relatively low temperature ($T$ $<$ 20 K) and high volume density ($n$(H$_2$) $>$ 10$^{4}$~\ccm) regions, CO molecules freeze out and adsorb onto dust grains has been reported \citep[e.g.,][]{Willacy_1998,Tafalla2002}.
In the case of CO freeze-out onto dust grains, low-$J$ lines of rare CO isotopologues such as \coeighteen\ or C$^{17}$O cannot trace the dense and cold gas \citep[e.g.,][]{Caselli_1999}.
This phenomenon, which is called CO freeze-out onto dust grains or depletion (hereafter CO depletion), is important for understanding the chemistry of molecular clouds because it not only affects the gas-phase chemistry but also promotes the surface reaction in dust grains.

CO depletion has been actively studied in nearby low-mass star-forming regions \cite[e.g.,][]{Willacy_1998,Caselli_1999,Kramer_1999,Bergin_2007}.
They have reported that CO depletion is frequently detected in starless cores at $T$ $<$ 20 K and $n$(H$_2$) $>$ 3 $\times$ $10^4$~\ccm, and the spatial scale is mostly comparable to the size of the molecular cloud core ($\sim$0.05~pc).
This is because the free fall timescale and the depletion timescale coincide at the typical density of molecular cloud cores ($\simeq$10$^4$~\ccm).

In recent years, a small number of sources with pc-scale CO depletion have been reported from mapping observations of the CO isotopologue lines \cite[][]{Hernandez_2011,Jimenez-Serra_2014,Feng_2016,Feng_2020,Gong_2018,Sabatini_2019,Sabatini_2022,Lewis_2021}.
In infrared dark clouds (IRDCs), which have higher densities than low-mass star-forming regions, pc-scale CO depletion is expected to be a common event; pc-scale CO depletion has been reported over a wide mass range, from low-mass IRDC, the Serpens filament \cite[20 -- 66 $\MO$,][]{Roccatagliata_2015,Gong_2018} to IRDCs that massive enough to form high-mass stars \cite[e.g.,][]{Sabatini_2019,Feng_2020}. 
In low-mass star-forming regions, CO depletion occurs in cloud cores, whereas in IRDCs CO depletion occurs not only in cloud cores but also in clumps and filaments. However, previous studies are biased towards high-mass IRDCs that fulfil an empirical threshold for high-mass star formation, $M(r)>870\ \MO$ (radius/pc)$^{1.33}$ \citep{Kauffmann_2010}. To better understand pc-scale CO depletion, more low-mass/intermediate-mass sources with pc-scale CO depletion are required.

\begin{figure*}
	\includegraphics[width=\linewidth, trim={20 310 15 5}, clip]{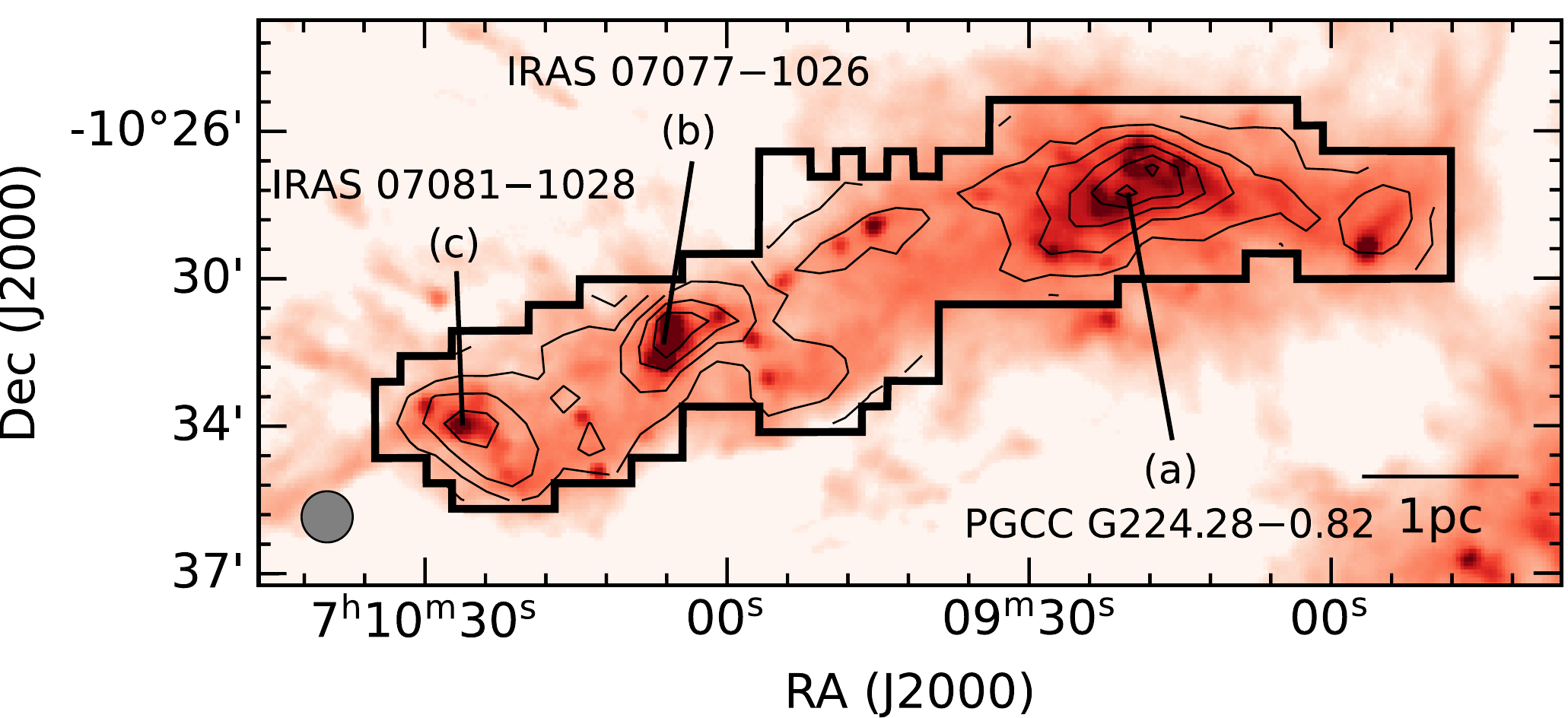}
    \caption{The \ammonia\ (1,1) integrated intensity map of KAG71 in contours over the  \textit{Herschel}/SPIRE 250 \micron\ dust continuum image. The integrated velocity range is from 11.9 km s$^{-1}$ to 19.0 km s$^{-1}$, containing only the main quadrupole hyperfine line. The black polygon outlines the mapping area of the \ammonia\ observations. The labels (a) -- (c)  indicate the positions for the spectra of figures~\ref{fig:spectra}-(a) to (c). The lowest contour and contour steps are 0.2 K~\kms\ and 0.4 K~\kms, respectively. The Nobeyama 45 m beam size at 23 GHz is indicated by the grey filled circle shown in the lower left corner. (Color online)}
    \label{fig:target_source}
\end{figure*}

\begin{figure*}
	\includegraphics[width=\linewidth, trim={55 0 93 40}, clip]{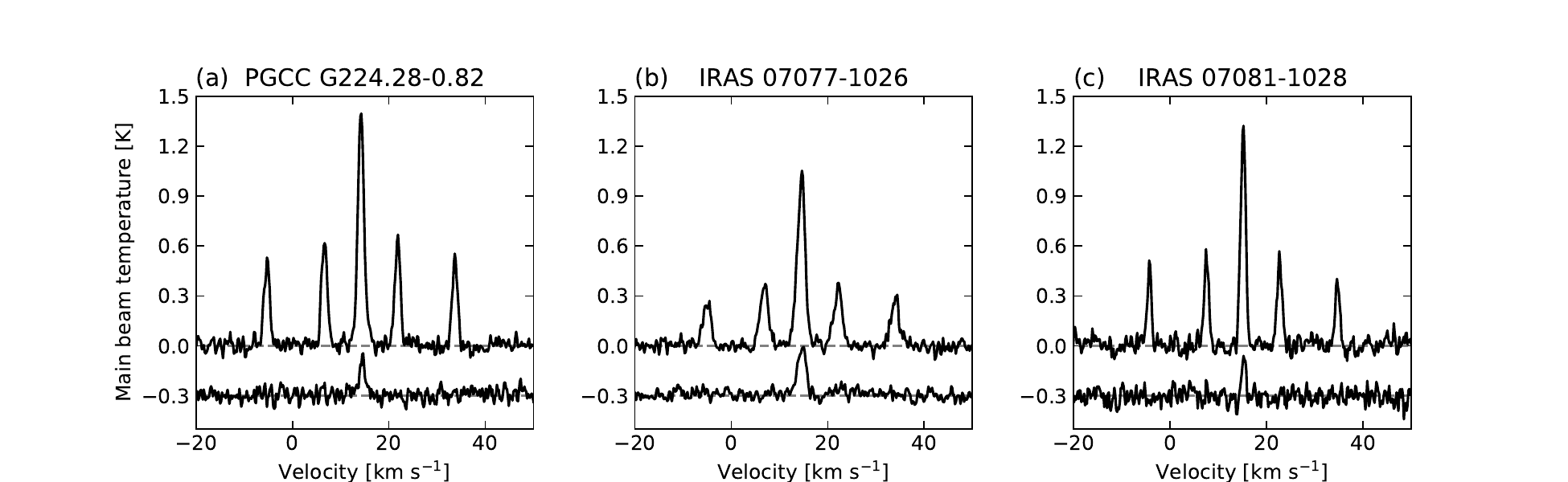}
        \caption{Panel (a) -- (c) : The spectra in the \ammonia\ (1,1) (top) and (2,2) (bottom) at the position assigned in figure~\ref{fig:target_source} with the same labels. (Color online)}
    \label{fig:spectra}
\end{figure*}

We focus on \ammonia\ molecular species. This molecule species is a good gas tracer in dense and cold regions, where star formation can occur \citep[e.g.,][]{Myers_1983}. The critical density of the \ammonia\ inversion transition lines are $\sim$2.0 $\times$ 10$^3$ \ccm\ \citep{Shirley_2015}, which is comparable to that of the \coeighteen\ ($J$=1--0) lines \citep[$\sim$2.0 $\times$ 10$^3$ \ccm,][]{Pety_2017}. These molecule lines are a good combination for comparing the properties of gases other than density differences. Previous studies of \ammonia\ lines reported that \ammonia\ is a depletion-resistant species and does not freeze out only in cold and dense gas regions ($T$ $<$ 20 K, $n \geq 10^5$ \ccm), where carbon-bearing molecules have already been frozen out onto dust grains \citep[e.g.,][]{Tafalla2002,Crapsi_2007,Sipila_2019}. Comparing the spatial distributions of \ammonia\ and \coeighteen\ may be able to detect pc-scale CO depletion. If this comparison can detect pc-scale CO depletion, it would be possible to separate the line-of-sight components, which is difficult to do with dust emission comparisons.

We observed \ammonia\ lines towards a star-forming region associated with the Canis Major (CMa) OB1 region. Figure~\ref{fig:target_source} shows the our observed area superimposed on the \textit{Herschel}\footnote{\textit{Herschel} is an ESA space observatory with science instruments provided by European-led Principal Investigator consortia and with important participation from NASA.} \citep{Pilbratt_2010} 250 \micron\ dust continuum image. This star-forming region is one of the targets of KAGONMA (Kagoshima Galactic Object survey with Nobeyama 45-metre telescope by Mapping in Ammonia lines) survey project \citep[][]{Murase_2021,Kohno_2022,Kohno_2023}, which is identified as KAGONMA 71 (hereafter KAG71). KAG71 corresponds to Canis Major Group 00 \citep{Fischer_2016} and Hi-GAL $l$ = 224$\degr$ region main filament \citep{Olmi_2016}. KAG71 contains Planck Catalogue of Galactic Cold Clumps \citep[PGCC:][]{Planck_XXVIII_2016} G224.28--0.82, IRAS 07077--1026, and IRAS 07081--1028.
\cite{Elia_2013} reported that the dust temperature, derived from the \textit{Herschel} dust continuum data, was between 11 K and 13 K in most of this region.
They also identified compact objects such as cores and clumps, and reported that most of the compact objects were associated with protostars. Previous studies have reported cluster formations \citep{Fischer_2016,Sewi_o_2019} and identified the outflows from protostars associated with clusters \citep{Sewi_o_2019,Lin_2021}. This suggests that our target is a very young and active star-forming region.
The distance to CMa OB1 is estimated by various methods and is still uncertain: 1150 pc based on the colour magnitude diagram \citep{Claria_1974}, 990 $\pm$ 50 pc based on $uvby\beta$ photometry \citep{Kaltcheva_2000}, and 1100 pc based on kinematic distances derived from $^{13}$CO ($J$=1--0) data \citep[][]{Kim_2004}.
There are several molecular clouds in the line-of-sight to CMa OB1. Their distances are individually estimated from kinematic distances using CO ($J$=1--0) data \citep{Elia_2013}, and a cloud corresponding to KAG71 is  at a distance of 900 pc. We adopt 900 pc to be consistent with previous studies \citep{Elia_2013,Sewi_o_2019}.

This paper is organised as follows: in section~\ref{sec:observations}, we describe the set-up of our observations, data reduction, and archival data sets. In section~\ref{sec:results}, we present the results of the spatial distribution of the \ammonia\ emission, the \coeighteen\ ($J$=1--0) emission and the dust continuum emission. The difference of spatial distributions is discussed and possible causes are described in section~\ref{sec:discussion}. In section~\ref{sec:conclutions}, we summarise our results and conclusions.

\begin{table*}
\caption{The HPBW and main-beam efficiency at each polarisation for each observation season.}
\centering
\begin{tabular}{ccccc}
\hline
season & HPBW (RHCP) & HPBW (LHCP) & $\eta_\mathrm{mb}$ (RHCP) & $\eta_\mathrm{mb}$ (LHCP) \\ \hline \hline
\begin{tabular}[c]{@{}c@{}}2015-2016\\ 2016-2017\end{tabular} & 74\arcsec.4 $\pm$ 0\arcsec.3 & 73\arcsec.9 $\pm$ 0\arcsec.3 & 83\% $\pm$ 4\% & 84\% $\pm$ 4\% \\
\begin{tabular}[c]{@{}c@{}}2017-2018\\ 2018-2019\end{tabular} & 73\arcsec $\pm$ 1\arcsec & 72\arcsec $\pm$ 1\arcsec & 84\% $\pm$ 2\% & 83\% $\pm$ 2\% \\
\hline
\end{tabular}
\label{tab:beam_eff}
\end{table*}

\section{Observations and data}
\label{sec:observations}

\subsection{\ammonia\ and H$_2$O Maser Observations}
\label{sec:NH3_H2O_obs}

We made mapping observations covering an area of 26\arcmin\ $\times$ 10\arcmin\ area (figure~\ref{fig:target_source}) with the Nobeyama 45 m radio telescope\footnote{The Nobeyama 45 m radio telescope is operated by the Nobeyama Radio Observatory, a branch of the National Astronomical Observatory of Japan.} from 2016 December to 2019 May. These observations used a high electron mobility transistor receiver, H22, and an auto-correlation spectrometer, the Spectral Analysis Machine for the 45 m telescope, SAM45 \citep{kuno_2011}. The metastable \ammonia\ inversion transitions at ($J,K$) = (1,1), (2,2), and (3,3), and H$_2$O maser were observed simultaneously in both circular polarisations. We observed 295 positions on a 37\arcsec.5 grid in the equatorial coordinates and the three-ON points position-switch observations. The half-power beam width (HPBW) varies slightly between the observing seasons and between the two circular polarisations (see table~\ref{tab:beam_eff}). We refer to 75\arcsec\ (corresponding to 0.33 pc at 900 pc), which is twice the grid size, as the effective beam size of the map instead of the HPBW. The OFF position was taken at ($\alpha_\mathrm{J2000},\delta_\mathrm{J2000}$) = (07:09:20.31, $-$10:27:55.35), where no \ammonia\ emission lines and no H$_2$O maser emission line were detected. The pointing accuracy was checked every hour using the H$_2$O maser sources associated with IK Tau, VY CMa, and Z Pup, and was within 7\arcsec. The rest frequencies \ammonia\ ($J,K$) = (1,1), (2,2), (3,3), and H$_2$O maser are 23.694495 GHz, 23.722633 GHz, 23.870129 GHz, and 22.235080 GHz, respectively. The bandwidth and frequency resolution are 62.5 MHz and 15.26 kHz, respectively, corresponding to 400 \kms\ and 0.19 \kms\ at the \ammonia\ (1,1) frequency. 
During the observations, the system noise temperature, $T_\mathrm{sys}$, that is the noise contributions from the atmosphere and the receiver, was between 100~K and 1000~K, or typically 200~K.
The antenna temperature ($T_\mathrm{a}^*$) was calibrated by the chopper wheel method. 
The observations at position (a) in figure~\ref{fig:target_source} were made at the beginning of each observation from December 2018 to March 2019. Twenty-eight observations were made, and the \ammonia\ (1,1) peak intensity had a variation of about 7\% at 1$\sigma$. The calibration accuracy of intensity is therefore $\sim$7\%.

Data reductions were performed using the Java NEWSTAR software package developed at the Nobeyama Radio Observatory (NRO). Baseline correction was performed with a third-order polynomial for emission-free channels. The main beam efficiency ($\eta_\mathrm{mb}$) at 23~GHz is different for each observation season and polarisation (see table~\ref{tab:beam_eff}). We used the mean value, 83.5\%, as $\eta_\mathrm{mb}$ for the whole observation season to convert from $T_\mathrm{a}^*$ to $T_\mathrm{mb}$. To check the consistency and reduce the noise level, we observed one or more times at the same position and averaged both polarisation data with a weight of 1/rms$^2$. To reduce the noise level, we also smoothed the \ammonia\ data to a velocity resolution of 0.38~\kms. The resulting rms noise level (hereafter $\sigma_\mathrm{rms}$) is typically 0.027~K on the $T_\mathrm{mb}$ scale. For the H$_2$O maser observations, a conversion factor of 2.8~Jy~K$^{-1}$ was used to convert the antenna temperature to flux density.

\subsection{Archival Data / Catalogue}

We used the \co\ ($J$=1--0) and \coeighteen\ ($J$=1--0) data cubes obtained with the Mopra 22~m telescope\footnote{\url{https://cdsarc.cds.unistra.fr/viz-bin/cat/J/A+A/594/A58}} \citep{Olmi_2016}. The beam size and velocity resolution are 38\arcsec\ and 0.09~\kms, respectively. We used the main beam efficiency of 0.42 \citep{Ladd_2005} measured at 115 GHz to convert the antenna temperature to the brightness temperature. Baseline correction was performed with a first-order polynomial for emission-free channels. To reduce the noise level, we smoothed over six spectral channels and the resulting velocity resolution is $\simeq$0.55 \kms. We have also produced regridded data with a grid size of 14\arcsec\ for comparison with the \textit{Herschel} dust continuum data. The mean rms noise of the 14\arcsec\ data is 0.39~K for \co\ and 0.18~K for \coeighteen\ on the $T_\mathrm{mb}$ scale, respectively. Although the observed area of the Mopra data does not cover the whole of our \ammonia\ observed area, the uncovered area is small; about 2\arcmin\ at the eastern and western edges of our area.

In order to understand the \coeighteen\ distribution in the entire \ammonia\ observed area, we also used the \coeighteen\ ($J$=1--0) data cube from FOREST Unbiased Galactic plane imaging survey \citep[FUGIN\footnote{ \url{http://jvo.nao.ac.jp/portal/nobeyama/fugin.do}}:][]{Umemoto_2017}, which is one of the legacy projects of the NRO. The FUGIN \co\ ($J$=1--0) data cube was also used to check the intensity consistency between the Mopra data and the FUGIN data. The effective beam size and velocity resolution are 20\arcsec\ and 1.3 \kms\ respectively. The mean rms noise is 3.6 K for \co\ and 1.5 K for \coeighteen\ on the $T_\mathrm{mb}$ scale, respectively. We smoothed the \coeighteen\ data to become its beam size of 75\arcsec\ and regridded it to a grid size of 37\arcsec.5 to match our \ammonia\ maps using the Astronomical Image Processing System (AIPS) software. The resulting mean rms noise is 0.25 K. 

The original FUGIN \coeighteen\ data are not sensitive enough, and the Mopra CO isotopologue lines data are smaller than our observed area. This makes it difficult to compare the intensity distributions of the \coeighteen\ line and the \ammonia\ line. In this paper, we used the smoothed FUGIN \coeighteen\ data to compare the line intensity distribution between \coeighteen\ and \ammonia, and the Mopra data for comparison with dust properties. The Mopra \co\ intensity is about 30\% higher than the FUGIN \co\ intensity, the \coeighteen\ intensities agree within 5\%. We discuss the effects of the different \co\ intensities on the physical parameters in appendix~\ref{sec:intensity_diff_12co}.

To understand the spatial distribution and physical conditions of the dust, we used the \textit{Herschel} dust continuum data from 160 $\micron$ to 500 $\micron$ with the identification numbers (OBSID) 1342220650 and 1342220651. These were obtained from the \textit{Herschel} Science Archive\footnote{\url{http://archives.esac.esa.int/hsa/whsa/}}. These data are a part of the \textit{Herschel} Infrared GALactic Plane Survey (Hi-GAL) key project \citep{Molinari_2010}. Different detectors were used for each wavelength range: for 160 \micron, the Photodetector Array Camera and Spectrometer \citep[PACS:][]{Poglitsch_2010}, and for 250 \micron, 350 \micron, and 500 \micron, the Spectral and Photometric Imaging REceiver \citep[SPIRE:][]{Griffin_2010}. We used the Level 2.5 data for each wavelength range.
The \textit{Herschel} PACS Level 2.5 calibrated data provide only relative photometry. We need to add a zero-level offset to the 160 \micron\ data. The offset calculation was performed by comparing \textit{Herschel} with dust properties data released by the \textit{Planck} collaboration \citep[][]{Planck_48_2016}, following the method of \cite{Lombardi_2014}. 
Because the \textit{Planck} data are at 5\arcmin\ resolution, we smoothed the \textit{Herschel} 160 \micron\ data to 5\arcmin\ resolution for comparison and derived the offset value of $-$73.7 MJy sr$^{-1}$ at 5\arcmin\ resolution. We added this value to the original resolution \textit{Herschel} data.
The beam of each wavelength is elongated along the scan direction during the observation. However, the elongation is small enough for discussion in this paper. Therefore, we used the geometric mean as the beam size. The beam size of four bands from 160 \micron\ to 500 \micron\ are 12\arcsec.6, 18\arcsec.4, 25\arcsec.2, and 36\arcsec.7, respectively\footnote{See details in photometer quick-start guide of each instrument. \url{https://www.cosmos.esa.int/web/herschel/legacy-documentation}}.

To study the impact of protostars on the interstellar medium, we used the catalogue of protostar candidates reported in \cite{Sewi_o_2019}. We only used YSO candidates for which the physical parameters were derived by spectral energy distribution (SED) fitting.

\section{Results}
\label{sec:results}
\subsection{Data overview}

\begin{figure}
	\includegraphics[width=\linewidth, trim={0 0 0 0}, clip]{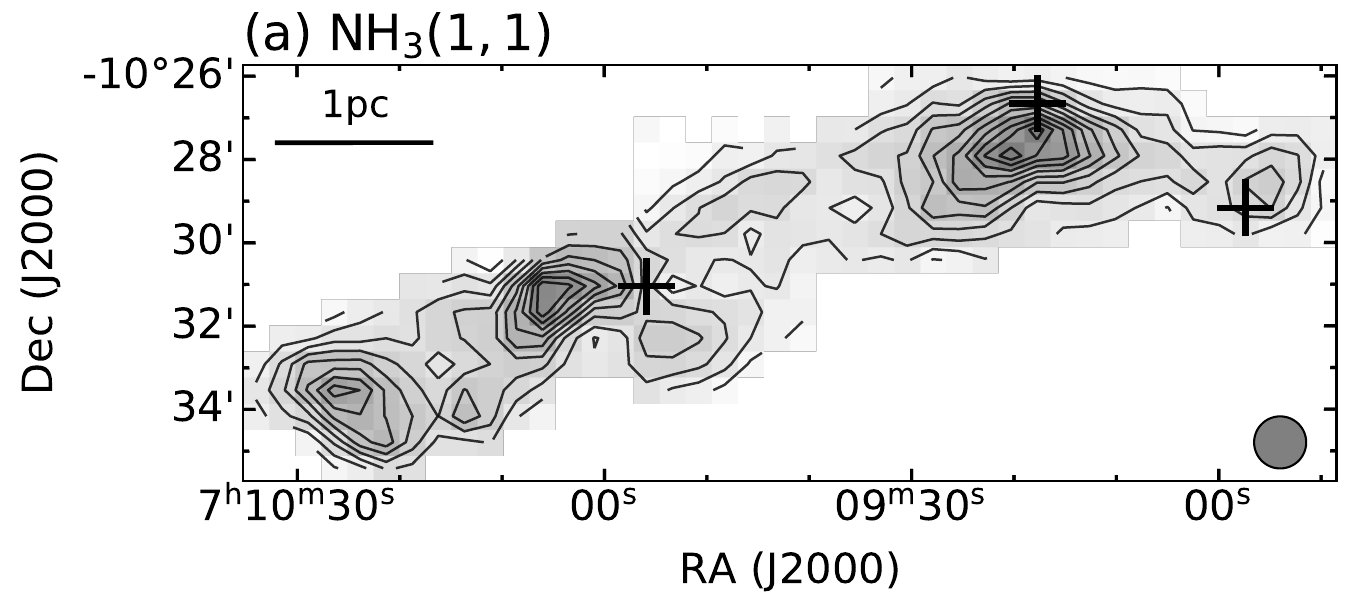}
	\includegraphics[width=\linewidth, trim={0 0 0 0}, clip]{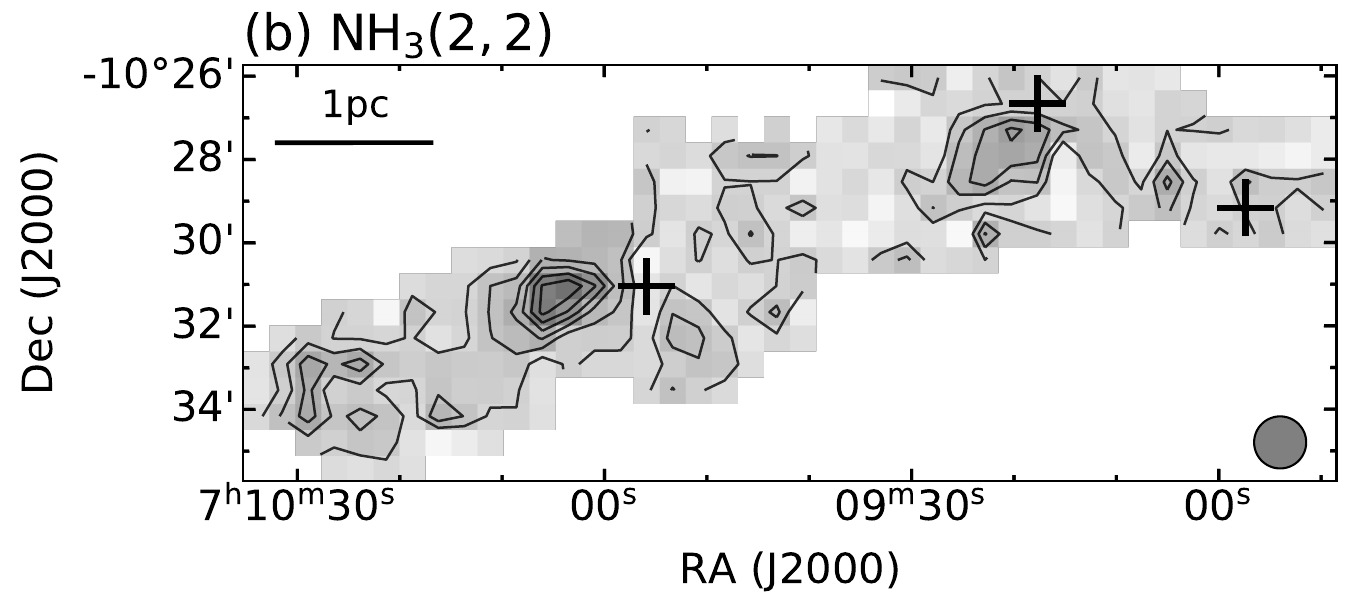}
    \caption{(a): Integrated intensity map of \ammonia\ (1,1). The lowest contour and contour steps are 0.2 K~\kms\ on the $T_\mathrm{mb}$ scale. Black crosses show the positions of the H$_2$O maser emission. (b): in (2,2). The lowest contour and contour steps are 0.1 K~\kms. In each panel, the integrated velocity range is from 11.8 \kms\ to 19.0 \kms. The beam size is indicated by the grey filled circle shown in the bottom right corner of each panel. (Color online)
    }
    \label{fig:integ_map}
\end{figure}

\begin{table*}
\caption{Coordinates and flux variations of H$_2$O masers.}
\centering
\begin{tabular}{cccccc}
\hline
ID & RA       & Dec       & 1st observation                                               & 2nd observation                                             & 3rd observation                                             \\ \hline \hline
1  & 07:08:57.4 & $-$10:29:10 & \begin{tabular}[c]{@{}c@{}}6.25 Jy\\ 03/12/2018\end{tabular} & \begin{tabular}[c]{@{}c@{}}6.12 Jy\\ 08/12/2018\end{tabular} & \begin{tabular}[c]{@{}c@{}}2.05 Jy\\ 16/02/2019\end{tabular} \\
2  & 07:09:17.7 & $-$10:26:40 & \begin{tabular}[c]{@{}c@{}}1.73 Jy\\ 14/02/2016\end{tabular}   &                                                             &                                                             \\
3  & 07:09:55.9 & $-$10:31:03 & \begin{tabular}[c]{@{}c@{}}0.85 Jy\\ 08/05/2017\end{tabular}   & \begin{tabular}[c]{@{}c@{}}0.76 Jy\\ 05/15/2017\end{tabular} &                                                             \\ \hline
\end{tabular}
\label{tab:water_maser}
\end{table*}

Figure~\ref{fig:spectra} shows the \ammonia\ (1,1) and (2,2) spectra at the position (a) -- (c) of figure~\ref{fig:target_source}. In our observations, we can find 5 quadrupole hyperfine lines consisting of one main quadrupole hyperfine line, two inner quadrupole satellite hyperfine lines, and two outer quadrupole satellite hyperfine lines in the (1,1) transition. In the (2,2) transitions only the main quadrupole hyperfine line was detected. Weak emission in the (3,3) main quadrupole hyperfine line ($T_\mathrm{mb} \simeq$ 0.11 K) was detected in the central position of IRAS 07077--1026 (position (b) shown in figure~\ref{fig:target_source}).

Figure~\ref{fig:integ_map}-(a) shows the integrated intensity map of the \ammonia\ (1,1) main quadrupole hyperfine line. We found three bright regions in the \ammonia\ emission associated with PGCC G224.28--0.82, IRAS 07077--1026, and IRAS 07081--1028. The spatial distribution of the \ammonia\ emission is consistent with the {\it Herschel} dust continuum emission. This suggests that \ammonia\ emission lines are good tracers of dense molecular gas. The detailed analysis is described in subsection~\ref{sec:Herschel} and intensity correlations between the \ammonia\ (1,1) emission and the 250 \micron\ dust emission are described in appendix~\ref{sec:correlations_NH3_Dust}. Although the \ammonia\ beam size ($\sim$0.3 pc) is larger than the typical filament width ($\sim$0.1 pc), the spatial distribution of the \ammonia\ emission is expected to reflect the filamentary structure derived from the dust emission reported by \cite{Schisano_2014} and \cite{Sewi_o_2019}.

Figure~\ref{fig:integ_map}-(b) shows the integrated intensity map of \ammonia\ (2,2) main quadrupole hyperfine line. The \ammonia\ (2,2) line was detected in the region where (1,1) was strongly detected, and the strongest emission was detected in IRAS07077--1026. This implies that IRAS 07077--1026 has a higher temperature than the other regions.

We detected three new H$_2$O maser sources. Table~\ref{tab:water_maser} shows the position, flux densities and observation dates of each maser source. Table~\ref{tab:water_maser} shows the centres of the highest intensity pixels. While IRAS 07077--1026 was reported to be associated with a H$_2$O maser \citep[e.g.,][]{Brand_1994,Sunada_2007}, it was not detected in our observations. Detected maser spectra are shown in figure~\ref{fig:maser_spectra}.

\subsection{Ammonia line fitting}

\label{sec:ammonia_fitting}

In case where the velocity dispersion $\gtrsim$ 0.2~\kms, \ammonia~(1,1) emission are observed only 5 quadrupole hyperfine lines. These quadrupole hyperfine lines which splited by electric quadrupole interactions, are further composed 18 magnetic hyperfine lines by magnetic spin-spin interactions. In this paper we refer to magnetic hyperfine lines as hyperfine components (HCs) and to quadrupole hyperfine lines as hyperfine component groups (HCGs).
These HCGs are traditionally called the main quadrupole hyperfine line, 2 inner quadrupole satellite hyperfine lines, and 2 outer quadrupole satellite hyperfine lines, which consist of 8, 3, and 2 HCs, respectively \citep[see][for details]{Ho_1977, Rydbeck_1977, Ho_1983,wang_2020}.

\begin{table*}
\caption{The parameters of hyperfine components for \ammonia\ (1,1).}
\centering
\begin{threeparttable}
\begin{tabular}{cccccc}
\hline
HCG&
\begin{tabular}{c}HC\\ number\end{tabular} & 
\begin{tabular}{c}F$_1^{\prime}$ $\rightarrow$ F$_1$\end{tabular} & 
\begin{tabular}{c}Relative\\ Intensity\tnote{a} \end{tabular} &
\begin{tabular}{c}fitting weight\\ {}$w_i$\tnote{b}{}\end{tabular}& 
\begin{tabular}{c}Velocity offset\\ {[}\kms{]}\end{tabular} \\ \hline \hline
Outer quadrupole satellite  & 1                & (1,0)                       & $\frac{2}{3}$                     & $\frac{2}{3}$                                                             & $-$19.55 \\
hyperfine line 1                                     & 2                & (1,0)                       & $\frac{1}{3}$                     & $\frac{1}{3}$                                                             & $-$19.41  \\ \hline
Inner quadrupole satellite  & 3                & (1,2)                        & $\frac{3}{5}$                     & $\frac{3}{5}$                                                             & $-$7.82  \\
hyperfine line 1                                     & 4                & (1,2)                        & $\frac{1}{15}$                  & $\frac{1}{15}$                                                            & $-$7.37 \\ 
                                     & 5                & (1,2)                        & $\frac{1}{3}$                    & $\frac{1}{3}$                                                              & $-$7.23 \\ \hline                                                                                 
Main quadrupole hyperfine line                   & 6               & (2,2)                        & $\frac{1}{25}$                  & $\frac{1}{30}$  & $-$0.25 \\
                                     & 7                & (1,1)                        & $\frac{5}{9}$                    & $\frac{5}{54}$  & $-$0.21 \\
                                     & 8                & (2,2)                        & $\frac{14}{25}$                & $\frac{7}{15}$  & $-$0.13 \\
                                     & 9                & (1,1)                        & $\frac{1}{9}$                    & $\frac{1}{54}$  & $-$0.07  \\
                                     & 10               & (2,2)                       & $\frac{9}{25}$                  & $\frac{3}{10}$   & 0.19  \\
                                     & 11               & (2,2)                        & $\frac{1}{25}$                  & $\frac{1}{30}$   & 0.31 \\
                                     & 12               & (1,1)                       & $\frac{1}{9}$                     & $\frac{1}{54}$   & 0.32 \\
                                     & 13               & (1,1)                       & $\frac{2}{9}$                     & $\frac{1}{27}$  & 0.46 \\ \hline
Inner quadrupole satellite & 14               & (2,1)                       & $\frac{1}{15}$                   & $\frac{1}{15}$                                                            & 7.35 \\
hyperfine line 2                                     & 15               & (2,1)                      & $\frac{3}{5}$                      & $\frac{3}{5}$                                                             & 7.47  \\
                                     & 16               & (2,1)                       & $\frac{1}{3}$                     & $\frac{1}{3}$                                                             & 7.89  \\ \hline
Outer quadrupole satellite & 17               & (0,1)                      & $\frac{2}{3}$                      & $\frac{2}{3}$                                                             & 19.32 \\
hyperfine line 2                                     & 18               & (0,1)                      & $\frac{1}{3}$                      & $\frac{1}{3}$                                                              & 19.85 \\ \hline
\end{tabular}
\begin{tablenotes}
\item[a] The relative intensities are taken from table 15 of \cite{Mangum_2015}. The sum of the relative intensities of each of magnetic hyperfine splittings associating with each quadrupole hyperfine splitting are 1.
\item[b] The intensity ratio of (F$_1^{\prime}$, F$_1$) = (1,1) to (2,2) is five to one; this ratio is used to correct for fitting weights of the main group.
\end{tablenotes}
\end{threeparttable}
\label{tab:fitting_param}
\end{table*}\textbf{}

To obtain the line width, we performed a relative intensity weighted Gaussian fit to all 18 HCs simultaneously at each observed position using the following equation \citep[similar fitting procedure is described in][]{Dhabal_2019}:
\begin{equation}
\label{eq:ammonia_fit}
    T_\mathrm{mb}(v)=\sum_{n=1}^{5} a_n \sum_{i} w_i \exp\left(-\frac{v-v_{\mathrm{los}}-\delta v_i}{2 \sigma_v^2}\right),
\end{equation}
where $a_n$ is the peak intensity of each HCG ($n$ = 1 to 5) in the case of well-blended HCGs are observed, $v_\mathrm{los}$ is the line-of-sight velocity of the main quadrupole hyperfine line, $\sigma_v$ is the velocity dispersion, $\delta v_i$ is the velocity offset of each HC from $v_\mathrm{los}$, and $w_i$ is the fitting weight of each HC \cite[$i$~=~1 to 18 calculated from tables 11 and 15 of ][]{Mangum_2015}. The $w_1$ to $w_5$ and $w_{14}$ to $w_{18}$ correspond to the relative hyperfine intensities of the magnetic hyperfine. \ammonia\ (1,1) main quadrupole hyperfine line contains two electric quadrupole hyperfine states of $\Delta F=0$. Therefore the $w_6$ to $w_{13}$ are calculated from the relative intensities of the magnetic hyperfine component and the electric quadrupole hyperfine component; the relative intensities of the magnetic hyperfine component are scaled by the relative intensities of the electric quadrupole hyperfine component and the sum of the $w_6$ to $w_{13}$ is 1. The $w_i$ and $\delta v$ are summarised in table~\ref{tab:fitting_param}. 
When the line width of the HCs is narrow, $a_n$ is not equal to the observed peak intensity of each HCG because individual HCs belonging to the same HCG do not overlap sufficiently, and the shape of the HCG does not follow a single Gaussian profile.
Since the line width obtained from our observations are broad, we assume that $a_n$ is equal to the observed peak intensity of each HCG. We assume that all 18 HCs have a same velocity dispersion, and each HC at the same HCG does not cause a hyperfine intensity anomaly.

We used a data cube with a velocity resolution smoothed to 0.38 \kms\ to improve the signal-to-noise ratio (S/N) of the quadrupole satellite hyperfine lines. Line fitting was performed at positions where the main quadrupole hyperfine line detected more than 5$\sigma_\mathrm{rms}$. When $\sigma_v$ $>$ 1.0 \kms, \ammonia\ main quadrupole hyperfine line and inner quadrupole satellite hyperfine lines begin to overlap \citep[][]{Zhou_2020}. We applied $\sigma_v$ under the condition of 0.2 \kms\ $<$ $\sigma_v$ $<$ 1.3 \kms\ to account for noise effects in these fits, because no overlapping profile between main and inner quadrupole satellite hyperfine lines is observed.

\begin{figure*}
        \includegraphics[width=\linewidth, trim={0 0 0 0}, clip]{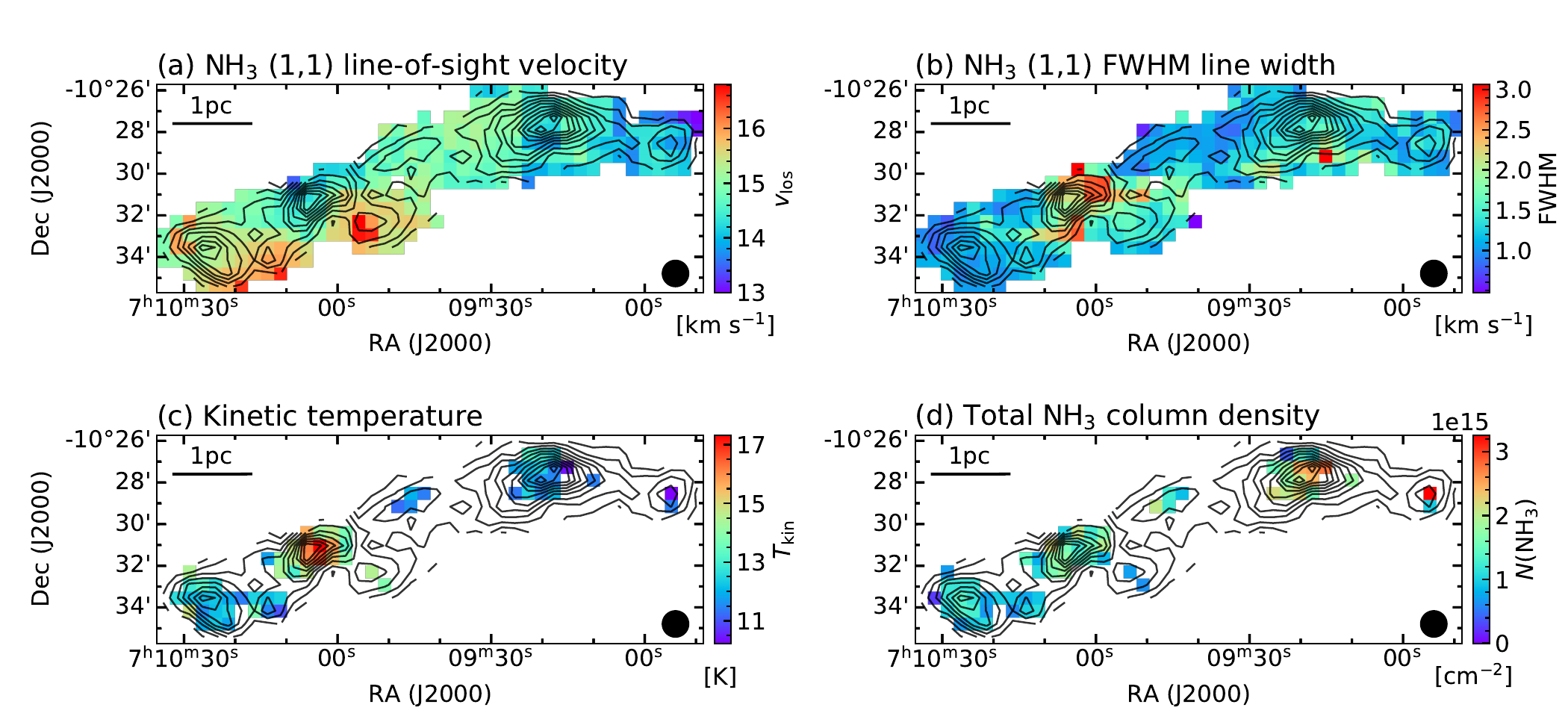}
    \caption{Spatial distributions of the physical parameters from \ammonia\ observations. (a): The line-of-sight velocity of the \ammonia\ (1,1) main quadrupole hyperfine line, (b): FWHM line width from the \ammonia\ (1,1) lines described in subsection~\ref{sec:ammonia_fitting}, (c): Kinetic temperature, and (d): Total \ammonia\ column density. The contours and the black filled circle in the bottom right corner are the same as in figure~\ref{fig:integ_map}-(a). (Color online)
    }
    \label{fig:nh3_params}
\end{figure*}

Figures~\ref{fig:nh3_params}-(a) and (b) show maps of $v_\mathrm{los}$ and full width at half maximum (FWHM) line widths estimated by \ammonia\ (1,1) fittings. We can see a velocity gradient from northwest to southeast. The FWHM ranges from 0.6 \kms\ to 3.1 \kms. The region of IRAS 07077--1026 shows that the line widths are broad.

In our observations, only the \ammonia\ (2,2) main quadrupole hyperfine line was detected. Although the main quadrupole hyperfine line consists of 12 HCs, single Gaussian fits were performed on the \ammonia\ (2,2) main quadrupole hyperfine line because the frequency of the 12 HCs in the main quadrupole hyperfine line is too close to resolve them in our observations. In these fits, the line widths of the \ammonia\ (1,1) and (2,2) lines were assumed to be the same \citep[see also][]{Urquhart_2011,Murase_2021}, and the fits were performed at positions where the \ammonia\ (2,2) main quadrupole hyperfine line was detected above 3$\sigma_\mathrm{rms}$. As with \ammonia\ (1,1), we used a data with a velocity resolution smoothed to 0.38 \kms.

The estimated \ammonia\ physical parameters were used in subsection~\ref{sec:physic_params}. 
For the line-of-sight velocity and velocity dispersion, the estimated covariances from these fits were taken as errors. For the peak intensity of the \ammonia\ (1,1) and (2,2) quadrupole hyperfine lines, the estimated covariances and the intensity calibration accuracy $\sim$7\% were taken as errors.

\subsection{Physical parameters from \ammonia\ lines}

\label{sec:physic_params}
We derived the optical depth, $\tau$, the kinetic temperature, $T_\mathrm{kin}$, and the column density, $N$(\ammonia), from our \ammonia\ observations. 

\subsubsection{optical depth $\tau$}

Under the local thermal equilibrium (LTE) conditions, the optical depth can be derived from the intensity ratio between the main quadrupole hyperfine line and quadrupole satellite hyperfine lines of the \ammonia\ (1,1) emission \citep{Ho_1983}. The optical depth can be derived from the following equation:
\begin{equation}
\label{eq:Ho_tau}
   \frac{T_{\mathrm{mb}}(\mathrm{sate})}{T_{\mathrm{mb}}(\mathrm{main})}
   =
   \frac{1-\exp(-a\tau_{(1,1,\mathrm{m})})}{1-\exp(-\tau_{(1,1,\mathrm{m})})},
\end{equation}
where $\tau_\mathrm{(1,1,m)}$ is the optical depth of the main quadrupole hyperfine line and $a$ is the relative intensity ratio of main quadrupole to quadrupole satellite hyperfine lines, which is 0.27778 for the inner quadrupole satellite hyperfine lines or 0.22222 for the outer quadrupole satellite hyperfine lines \citep{Mangum_1992}. In the optically thin case, the $a$ is the ratio of the total intensity of eight HCs forming the main quadrupole hyperfine line to the total intensity  of two or three HCs forming the quadrupole satellite hyperfine line \citep[see][]{Mangum_2015}. When the line width is relatively broad, the HCs are well-blended and each HCG has a single Gaussian profile, so the $a$ is approximately equal to the peak intensity ratio of the main quadrupole hyperfine line to the inner or outer quadrupole satellite hyperfine line. However, when the line width is narrow, the internal HCs are separated at each HCG and the $a$ does not match the peak intensity ratio between the main quadrupole hyperfine line and the inner or outer quadrupole satellite hyperfine line \citep[see][]{Zhou_2020,wang_2020}. Taking this effect into account, we derived the optical depth from the integrated intensity of the HCGs,
\begin{equation}
\label{eq:cal_tau}
   \frac{\int T_{\mathrm{mb}}(\mathrm{sate})\mathrm{d}v}{\int T_{\mathrm{mb}}(\mathrm{main})\mathrm{d}v}=\frac{1-\exp(-a\tau_{(1,1,\mathrm{m})})}{1-\exp(-\tau_{(1,1,\mathrm{m})})}.
\end{equation}
To improve the S/N, we used the mean integrated intensity of 2 inner quadrupole satellite hyperfine lines for the quadrupole satellite hyperfine line intensity. The optical depth was calculated for pixels where the peak intensity of both inner quadrupole satellite hyperfine lines is greater than 5$\sigma_\mathrm{rms}$.
The integrated velocity range for each HCG is approximately $v_{\mathrm{los}}$ $\pm$ 2.6$\sigma_v$. In this range, it is possible to integrate 99\% of the radiation represented by the Gaussian profile. As a result, the range of $\tau_{(1,1,\mathrm{m})}$ is from 0.3 to 2.2, and the mean value is 0.9.

\subsubsection{The gas temperature}

We used the following equation to derive the rotational temperature $T_\mathrm{rot}$ for pixels where the optical depth can be estimated and \ammonia\ (2,2) main quadrupole hyperfine line was detected above 3$\sigma_\mathrm{rms}$,
\begin{multline}
\label{eq:trot}
    T_{\mathrm{rot}}(2,2;1,1) =
    -41.2
    \Big/ \ln \Big( \frac{-0.282}{\tau_{(1,1,\mathrm{m})}} \\
    \times \ln \left[
    1-\frac{T_\mathrm{mb}(2,2,\mathrm{m})}{T_\mathrm{mb}(1,1,\mathrm{m})}
    \times
    [1-\exp(-\tau_{(1,1,\mathrm{m})})]
    \right] \Big) .
\end{multline}

We converted the rotational temperature to the kinetic temperature using the following equation \citep{Swift_2005}: 
\begin{equation}
    T_{\mathrm{rot}} = T_{\mathrm{kin}}\left\{1+\frac{T_{\mathrm{kin}}}{T_0}\ln[1 + 0.6 \times \exp(-15.7/T_{\mathrm{kin}})]\right\}^{-1},
\end{equation}
where $T_0$ is the energy difference between the (1,1) and (2,2) levels in Kelvin, which is 41.2 K. In the above equation, we used the value of a collision rate between \ammonia\ and hydrogen molecules reported by \cite{Danby_1988}.

Figure~\ref{fig:nh3_params}-(c) shows the spatial distribution of $T_\mathrm{kin}$. The range of $T_\mathrm{kin}$ is from 10.2 K to 17.3 K, and the mean value is 12.9 K. We found that the temperature around IRAS 07077--1026 is higher than the other regions. 

\subsubsection{The column density of \ammonia}

Under LTE conditions, \ammonia\ (1,1) column density, $N\mathrm{(1,1)}$, can be derived by the following equation \citep{Mangum_1992}:
\begin{equation}
    N(1,1) = 6.60 \times 10^{14} \frac{T_\mathrm{rot}}{\nu(1,1)} \tau_{(1,1,\mathrm{m})} \Delta V_{1/2},
\end{equation}
where $\nu(1,1)$ is the rest frequency of \ammonia\ (1,1) in GHz, and $\Delta V_{1/2}$ is the FWHM of the \ammonia\ (1,1) main quadrupole hyperfine line in \kms. The total \ammonia\ column density, $N$(\ammonia), is given by the following equation \citep{Mangum_1992}:
\begin{multline}
\label{eq:Ntot}
    N({\mathrm{NH_3}}) = \\
    N(1,1) \sum_J \sum_K \left[ 
    \frac{2g_J g_I g_K}{3}
    \exp \left( \frac{23.3-E(J,K)/k_B}{T_{\mathrm{rot}}} \right)
    \right],
\end{multline}
where $g_J$ is the rotational degeneracy, $g_I$ is the nuclear spin degeneracy, and $g_K$ is the $K$--degeneracy. $E(J,K)/k_\mathrm{B}$ is the energy difference (in Kelvin) of each excited state from the ground state, using data from the JPL Molecular Spectroscopy catalogue \citep{Pickett_1998}. The total \ammonia\ column density was calculated up to metastable states of ($J$,$K$) = (6,6), assuming that an ortho-to-para ratio is 1. The column density of \ammonia\ was calculated at positions where the rotational temperature could be derived as described above.

Figure~\ref{fig:nh3_params}-(d) shows that the spatial distribution of $N$(\ammonia) does not vary significantly in most of the regions. This figure shows that $N$(\ammonia) is distributed in the range of $2 \times 10^{14}$ \scm\ to $3.3 \times 10^{15}$ \scm, and the mean value is $1.4 \times 10^{15}$ \scm.

\subsubsection{Error estimation}
Estimating the error in observed physical parameters (i.e., optical depth, gas temperature, and column density) is important. In this paper, we evaluated the errors in these parameters by Monte Carlo method for each pixel \citep[][see for a similar algorithm]{Murase_2021}. There are three following steps in error estimation, and we applied these in turn to equations~(\ref{eq:cal_tau})--(\ref{eq:Ntot}):

\begin{enumerate}
    \item consider the known error in the physical parameter as a standard deviation and generate a random number that follows a normal distribution;
    \item substitute the best estimate of the known physical parameter + random number into the equation and derive the physical parameter with an unknown error. In the case of estimating the physical parameter using line intensity, multiply the observed value by the random number that follows the flux calibration error;
    \item repeat the above 10000 times and derive the standard deviation from the physical parameters obtained and consider it as the error inherent in the physical parameter.
\end{enumerate}

The estimated errors of $T_\mathrm{kin}$ and $N$(\ammonia) for \ammonia\ bright positions are summarised in table~\ref{tab:dendro}.

\subsection{Physical parameters from CO lines}
\label{sec:phy_param_c18o}
As a first step to figure out the physical conditions of the gas traced by the \coeighteen\ emission, we derived the optical depth, $\tau$, the excitation temperature, $T_\mathrm{ex}$, and the column density, $N$(\coeighteen), derived from the Mopra data. Before estimating these parameters, we performed single Gaussian fits to the \coeighteen\ line. In this paper, we only used spectra with a S/N $\geq$ 3 to estimate the physical parameters. 

Assuming that the \co\ emission is optically thick and the beam-filling factor is unity, the excitation temperature can be estimated by the following equation \citep[e.g.,][]{Nagahama_1998}:
\begin{equation}
\label{eq:co_tex}
    T_\mathrm{ex}=5.53
    \left[ \ln 
    \left( 1+\frac{5.53}{T_\mathrm{peak}(^{12}\mathrm{CO})+0.819} \right)
    \right]^{-1},
\end{equation}
where $T_\mathrm{peak}$(\co) is the peak intensity of the \co\ emission. The $T_\mathrm{ex}$ ranges from 10.5 K to 17.8 K and the mean value is 13.9 K. 

Assuming that the molecular clouds are under the LTE and have the same excitation temperature along the line-of-sight, and the beam-filling factor is unity, the optical depth of \coeighteen\ can be obtained from the equation \citep{Shimajiri_2014}:

\begin{equation}
    \label{eq:co_tau}
    \tau_\mathrm{C^{18}O} = 
    -\ln \left(
    1-\frac{T_{\mathrm{peak}}(\mathrm{C^{18}O})}{5.27([\exp(5.27/T_{\mathrm{ex}})-1]^{-1}-0.1666)}\right),
\end{equation}
where $T_\mathrm{peak}$(\coeighteen) is the peak intensity of the \coeighteen\ emission. The $\tau_\mathrm{C^{18}O}$ ranges from 0.04 to 0.44 and the mean value is 0.15. The assumption of the beam filling factor=1 underestimates the optical depth, which also affects later estimates of the column density; the extent to which it may underestimate the optical depth is discussed in subsection \ref{sec:tau_uncertainty}.

\begin{figure}
	\includegraphics[width=\linewidth, trim={5 5 5 5}, clip]{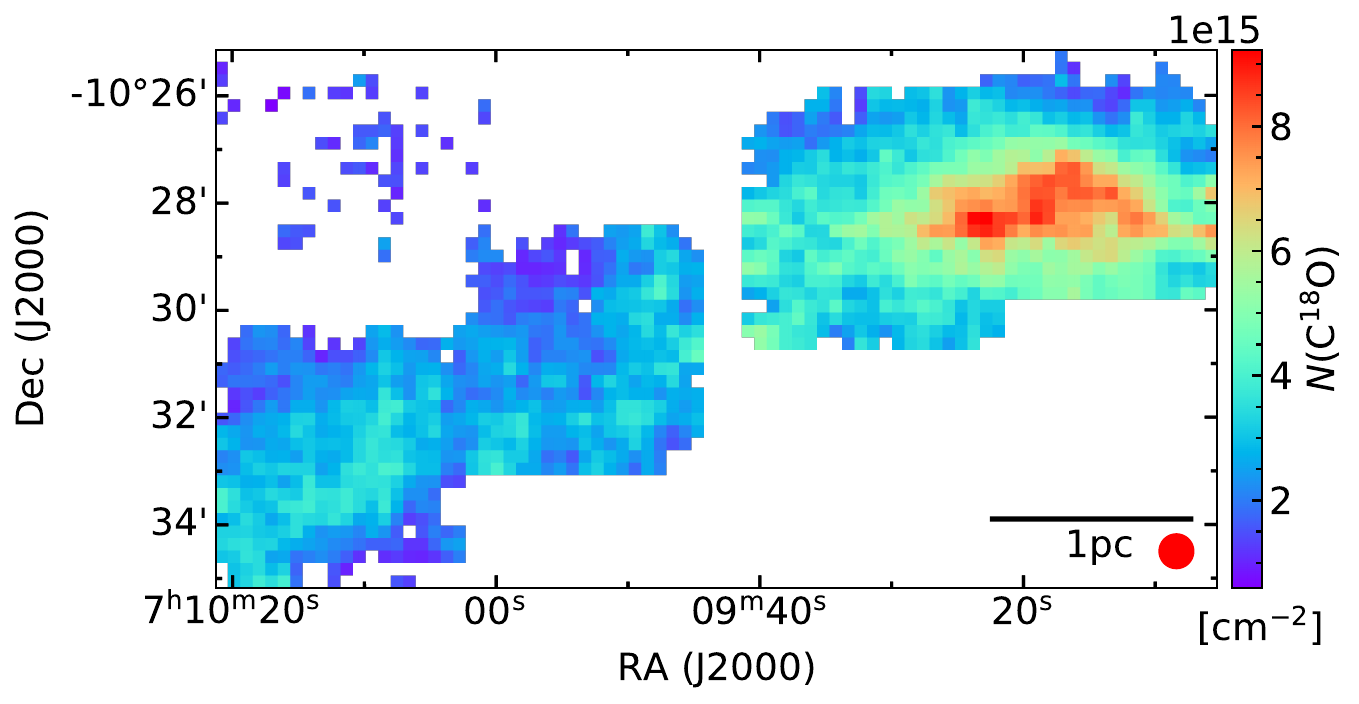}
    \caption{\coeighteen\ column density map from the Mopra data. The red filled circle in the bottom right corner indicates the beam size of the Mopra data. (Color online)}
    \label{fig:C18O_column}
\end{figure}

The column density of \coeighteen\ can be derived as follows \citep[see][for details]{Mangum_2015}:
\begin{equation}
\label{eq:co_column_density}
\begin{split}
    N(\mathrm{C^{18}O}) &= \frac{3h}{8\pi^3 S\mu^2} \frac{Q_\mathrm{rot}}{g_u}
    \frac{\exp\left(\frac{E}{\kappa_B T_\mathrm{ex}}\right)}{\exp\left(\frac{h\nu}{\kappa_B T_\mathrm{ex}}\right)-1}
    \int\tau dv\\
    &\simeq\frac{3h}{8\pi^3\mu^2} \frac{\exp\left(\frac{5.27}{T_\mathrm{ex}}\right)}{\exp\left(\frac{5.27}{T_\mathrm{ex}}\right)-1}
    \frac{Q_\mathrm{rot}}{J_\nu(T_{\mathrm{ex}})-J_\nu(T_{\mathrm{bg}})} \\
    & \times \frac{\tau_\mathrm{C^{18}O}}{1-\exp(-\tau_\mathrm{C^{18}O})}
    \int T_\mathrm{mb} dv,
\end{split}
\end{equation}
where $S$ = $J/(2J+1)$ is the line strength, $\mu$ is the electric dipole moment, which is 0.11079 Debye for \coeighteen, $g_u$ = $2J+1$ is the level degeneracy, $E/\kappa_B$ = 5.27 K is the energy difference of excited state from the ground state in Kelvin, $J_\nu(T)$ = $\frac{h\nu/k_B}{\exp(h\nu/(k_BT))-1}$ is the radiation temperature, and $T_\mathrm{bg}$ = 2.725 K is the temperature of cosmic microwave background radiation. The rest frequency of the \coeighteen\ ($J$=1--0) emission line, $\nu$, is 109.782182 GHz. 
Also, the rotational partition function, $Q_\mathrm{rot}$, is given by, 

\begin{equation}
\label{eq:co_partition_func}
Q_\mathrm{rot} = \sum_{J=0}^{\infty}(2J+1)\exp\left(-\frac{hB_0J(J+1)}{k_BT_\mathrm{ex}}\right),
\end{equation}
where $B_0$ is the rigid rotor rotation constant, which was set to 54891.4 MHz. The electric dipole moment and rigid rotor rotation constant were taken from the JPL molecular spectroscopy catalogue \citep{Pickett_1998}. The rotational partition function was calculated up to $J$=14. 
The errors of all physical parameters are derived by the Monte Carlo method. In these calculations, a flux calibration error of 7\% was used as the error for the chopper wheel method \citep{Ulich_1976}.

Figure~\ref{fig:C18O_column} shows the spatial distribution of the \coeighteen\ column density. It shows that \coeighteen\ molecular gases are concentrated at PGCC G224.28--0.82 and that there is no significant feature in the eastern part of the observed area. The $N$(\coeighteen) ranges from 0.6 $\times$ 10$^{15}$ \scm\ to 9.2 $\times$ 10$^{15}$ \scm\ and the mean value of $N$(\coeighteen) is 3.4 $\times$ 10$^{15}$ \scm. 

\subsection{Deriving dust properties and spatial distributions}
\label{sec:Herschel}

\begin{figure*}
    \includegraphics[width=\linewidth, trim={0 0 0 0}, clip]{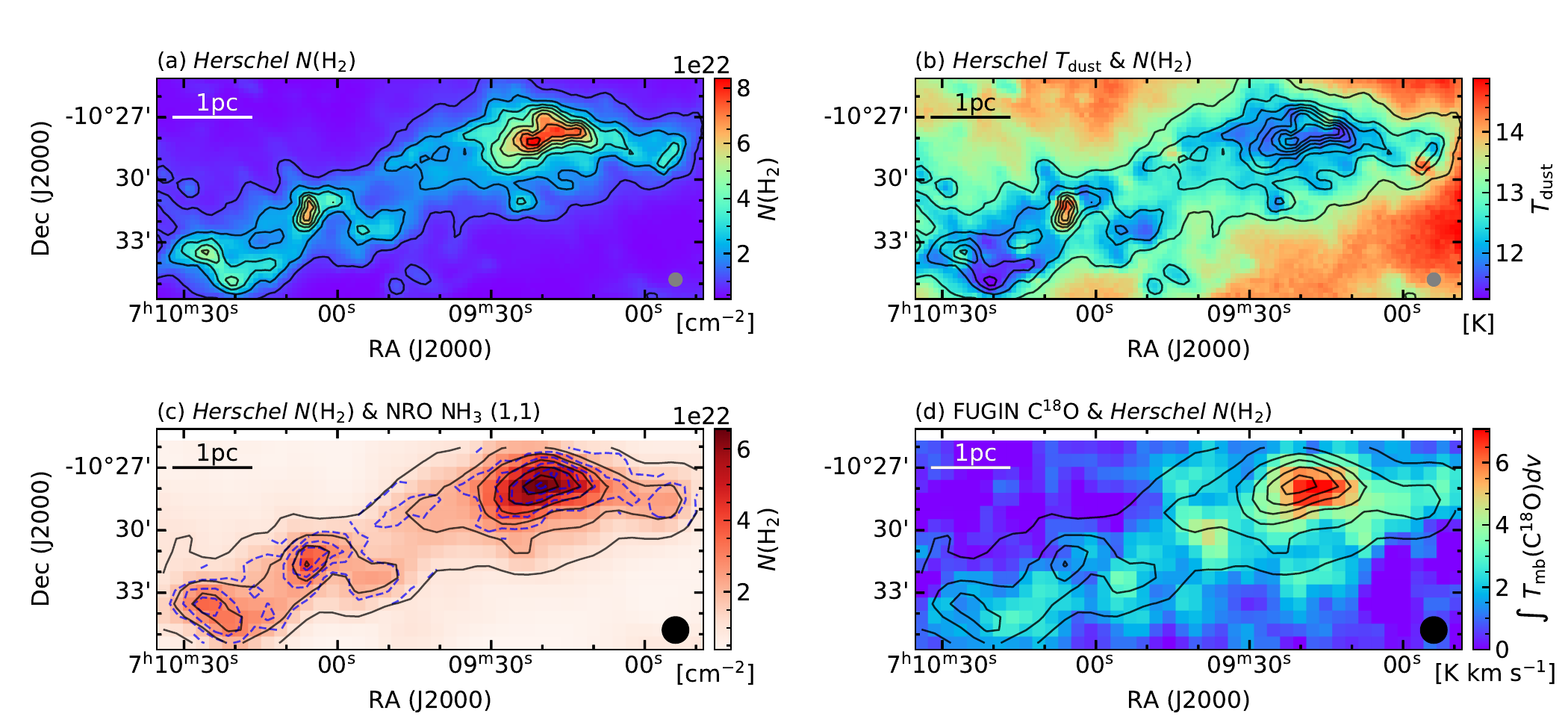}
    \caption{(a): The colour image and contours indicate the $N$(H$_2$) map obtained in subsection~\ref{sec:Herschel}. The lowest contour and contour steps are 1.0 $\times$ 10$^{22}$ \scm.
    (b): The dust temperature image obtained in subsection~\ref{sec:Herschel}. The contours are the same as that shown in panel (a). 
    (c): The colour image and black contours indicate the $N$(H$_2$) map derived from the \textit{Herschel} data smoothed to be the same beam size as the \ammonia\ data. The lowest contour and contour steps are 1.0 $\times$ 10$^{22}$ \scm.
    The blue dashed contours indicate \ammonia\ (1,1) integrated intensity, which are the same as in figure~\ref{fig:target_source}.
    (d): $N$(H$_2$) distributions (contours) superposed on the \coeighteen\ integrated intensity map with the velocity range from 11.7 \kms\ to 18.9 \kms\ (colour image). The black contours are the same as the black contours in panel (c).
    In all panels, grey filled circle indicates the beam sizes of the \textit{Herschel} data, and black filled circle indicates the 75\arcsec\ beam size. (Color online)}
\label{fig:Herschels}
\end{figure*} 

Using the $Herschel$ dust continuum data, we derived the dust temperature $T_\mathrm{dust}$ and the molecular hydrogen column density $N$(H$_2$), using the wavelengths, $\lambda$ from 160 \micron\ to 500 \micron. We made two data sets, the first is that all wavelength maps were smoothed to the beam size of 36.\arcsec7 and regridded to 14\arcsec, the second is that all wavelength maps were smoothed to the beam size of 75\arcsec\ and regridded to 37.\arcsec5.
Assuming that a molecular cloud has a single dust temperature along the line-of-sight and that the atomic hydrogen column density is negligibly low, SED fitting towards each pixel was performed by using the following equation: 
\begin{equation}
\label{eq:modified_bb}
I_\nu = B_\nu(T_\mathrm{dust})[1-e^{-\tau_\nu}], 
\end{equation}
where $I_\nu$ is the intensity at frequency $\nu$, $B_\nu$($T_\mathrm{dust}$) is the Planck blackbody function, $\tau_\nu = \kappa_\nu\Sigma$ is the dust optical depth, $\kappa_\nu$ is the dust opacity per unit mass (dust + gas), $\Sigma = \mu m_\mathrm{H} N$(H$_2$) is the surface gas density, $\mu$ is the mean molecular weight, and $m_\mathrm{H}$ is the atomic hydrogen mass. For consistency with \cite{Elia_2013}, we adopt $\kappa_\nu$ = $\kappa_\lambda$ = $0.1 \times (\frac{\lambda}{250 [\micron]})^{-\beta}$ [\scm\ g$^{-1}$] assuming a gas-to-dust mass ratio of 100, a dust emissivity index $\beta$ of 2 \citep{Hildebrand_1983}, and $\mu$ = 2.8 to take into account a relative helium abundance of 25\% by mass \citep{Kauffmann_2008}. Following \cite{Konyves_2015}, each data point of the SED fit was weighted by 1/$\sigma_\mathrm{cal}^2$, where $\sigma_\mathrm{cal}$ is the calibration error of surface brightness (20\% at 160 \micron\ and 10\% at 250 \micron, 350 \micron, and 500 \micron). Its error estimate uses the covariance matrix. 

Figure~\ref{fig:Herschels}-(a) shows the spatial distribution of the $N(\rm{H_2})$. The high-column density is found around PGCC G224.28--0.82, IRAS 07077--1026, and IRAS 07081--1028. The $N$(H$_2$) ranges from 0.2 $\times$ 10$^{22}$ to 8.3 $\times$ 10$^{22}$ \scm. 

Figure~\ref{fig:Herschels}-(b) shows the spatial distribution of the $T_\mathrm{dust}$. We can see that the $T_\mathrm{dust}$ is higher at the edge of the cloud and lower at the centre. This temperature structure has generally been observed in the nearby molecular clouds and infrared dark clouds \citep[e.g.,][]{Planck_2011_XXV,sokolov_2017}. The dust temperature in the \ammonia\ observed area is 11 K -- 14.5 K, which is almost the same as the distribution of the $T_\mathrm{kin}$ (figure~\ref{fig:nh3_params}-(c)). 

Figure~\ref{fig:Herschels}-(c) shows the map of the \ammonia\ (1,1) integrated intensity superposed on the $N$(H$_2$) derived from the dust continuum emission. In the \ammonia\ observed area, the \ammonia\ (1,1) integrated intensity and $N$(H$_2$) show good agreement with each other. On the other hand, the region between PGCC G224.28--0.82 and IRAS 07077--1026  has $N\mathrm{(H_2)} > 2.0 \times 10^{22}$ \scm, while the \ammonia\ emission is very weak.

Figure~\ref{fig:Herschels}-(d) shows the $N$(H$_2$) superposed on the map of the FUGIN \coeighteen\ ($J$=1--0) integrated intensity. We can find no significant correlation between the \coeighteen\ emission and the spatial distribution of $N$(H$_2$), except for PGCC G224.28--0.82. The \coeighteen\ emissions were detected in the region between PGCC G224.28--0.82 and IRAS 07077--1026.

\subsection{Identification of dense gas clumps with dendrogram} 
\label{sec:dendrogram_analysis}

\begin{table*}
\caption{Physical properties of the \ammonia\ clumps identified by the dendrogram.}
\centering
\begin{tabular}{cccccccc}
\hline
clump ID& RA & Dec & \begin{tabular}[c]{@{}c@{}}$R$ \\ {[}pc{]}\end{tabular} & \begin{tabular}[c]{@{}c@{}}$T_\mathrm{kin}$ \\ {[}K{]}\end{tabular} & \begin{tabular}[c]{@{}c@{}}FWHM \\ {[}\kms{]}\end{tabular} & \begin{tabular}[c]{@{}c@{}}$N$(\ammonia)\\ {[}\scm{]}\end{tabular} & \begin{tabular}[c]{@{}c@{}}Mass \\ {[}$\MO${]}\end{tabular} \\ \hline \hline
1     & 07:08:55.9 & $-$10:28:40.5 & 0.29 & 10.2 (0.7)& 1.32 (0.03) & 3.3 $\times$ 10$^{15}$ (6 $\times$ 10$^{14}$) & 90 (10) \\
2     & 07:09:19.3 & $-$10:27:42.3 & 0.32 & 11.6 (0.5)& 1.43 (0.01) & 2.1 $\times$ 10$^{15}$ (2 $\times$ 10$^{14}$)& 290 (30) \\
3     & 07:09:51.5 & $-$10:32:17.8 & 0.23 & 15 (1) & 1.64 (0.05)& 7 $\times$ 10$^{14}$ (5 $\times$ 10$^{14}$)& 54 (7) \\
4     & 07:10:05.4 & $-$10:31:28.2 & 0.26 & 16.1 (0.8)& 2.18 (0.02)& 1.6 $\times$ 10$^{15}$ (3 $\times$ 10$^{14}$)& 110 (10) \\
5     & 07:10:12.7 & $-$10:33:52.6 & 0.19 & 11.5 (0.8)& 1.26 (0.02)& 1.0 $\times$ 10$^{15}$ (3 $\times$ 10$^{14}$)& 41 (6) \\
6     & 07:10:24.4 & $-$10:33:47.6 & 0.39 & 12.1 (0.7)& 1.16 (0.02)& 1.4 $\times$ 10$^{15}$ (2 $\times$ 10$^{14}$)& 220 (30) \\ \hline
\end{tabular}
\begin{tablenotes}[]
\footnotesize 
        \item \textit{Notes:} The physical parameters shown in this table are derived from the positions where the strongest \ammonia\ (1,1) integrated intensities were obtained for each clump. The errors are shown in parentheses.
\end{tablenotes}
\label{tab:dendro}
\end{table*}

\begin{figure*}
	\includegraphics[width=\linewidth]{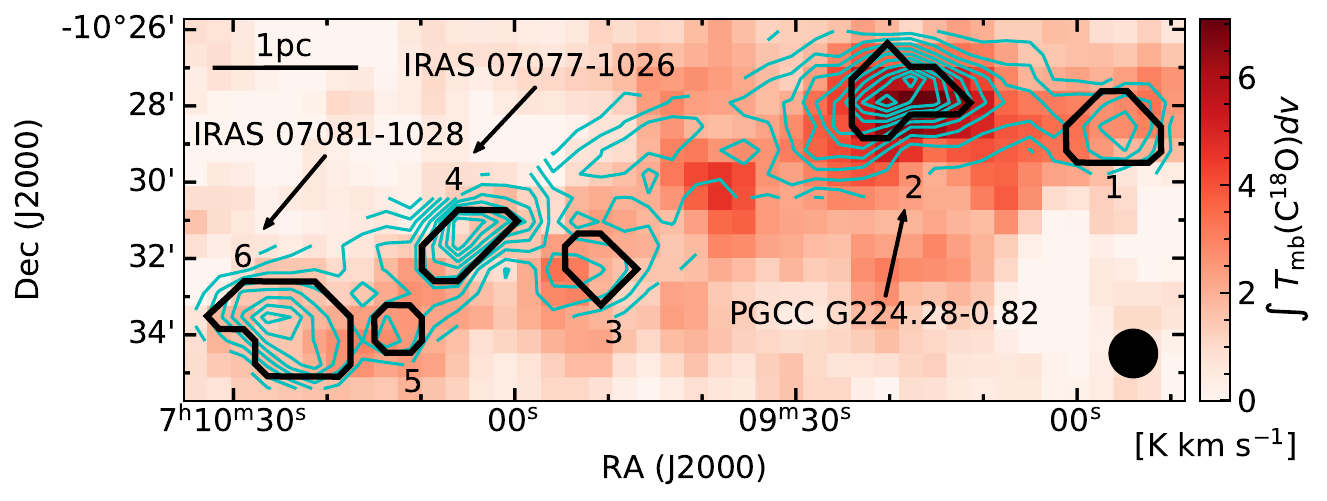}
    \caption{Integrated intensity map of \ammonia\ (1,1) in contours over the FUGIN \coeighteen\ ($J$=1--0) integrated intensity image. The thick black polygons indicate the position and size of the clumps obtained from the dendrogram analysis in subsection~\ref{sec:dendrogram_analysis}. The contour steps and the black filled circle in the bottom right corner are the same as figure~\ref{fig:integ_map}-(a). (Reprinted with permission from \cite{Handa_2023}. Copyright The Author(s), 2023 Published by Cambridge University Press on behalf of International Astronomical Union.) (Color online)
    }
    \label{fig:NH3_CO_integ_add_dendro}
\end{figure*}

To identify the hierarchical structure of the molecular cloud, we performed a dendrogram analysis \citep{Rosolowsky_2008} on the \ammonia\ (1,1) main quadrupole hyperfine line data cube, which has a velocity resolution of 0.19 \kms\ and a typical rms noise level of 0.035 K. This analysis uses the following three parameters: min\_value, min\_delta, and min\_npix, where min\_value is a parameter to distinguish between structure and noise, min\_delta is the threshold to separate multiple structures, and min\_npix is the minimum number of pixels to identify a structure, respectively. In this paper, we used the input with the min\_value = 3$\sigma$, min\_delta = 3$\sigma$, where $\sigma =$ 0.035 K is the rms noise, and min\_npix = 4 pixels, which is a minimum integer number of pixels larger than the beam size. In addition, we have added the condition that the peak intensity of the structure is greater than 7$\sigma$ and that the structure has a spread of two channels or more in the velocity axis. We refer to a "leaf" as a clump in the dendrogram output. 

Figure~\ref{fig:NH3_CO_integ_add_dendro} shows the results of the dendrogram analysis. The six \ammonia\ clumps were identified. We gave an identification number from west to east for each clump. 
The centre of PGCC G224.28--0.82 corresponds to the clump 2, IRAS 07077--1026 and IRAS 07081--1028 correspond to the clumps 4 and 6, respectively.
We adopted $R = \sqrt{A/\pi}$ as the clump radius, where $A$ is the total area on the sky in a clump. The clump radius ranges from 0.19 pc to 0.39 pc. We estimated the mass of each clump using Mass = $\mu m_\mathrm{H} A_\mathrm{pixel} \sum_{i} N_i(\mathrm{H_2})$, where $A_\mathrm{pixel}$ is the surface area of the pixel and the $N_i$(H$_2$) is the $N$(H$_2$) of each pixel belonging to each clump.  This calculation uses the $N$(H$_2$) data derived in subsection~\ref{sec:Herschel}, which has a resolution of 75\arcsec. The clump mass ranges from 41 $\MO$ to 290 $\MO$. Only the clump 2 is massive enough to fulfil the empirical threshold for high-mass star formation, $M(r)>870\ \MO$ (radius/pc)$^{1.33}$ \citep{Kauffmann_2010}. The physical parameters of the clumps are summarised in table~\ref{tab:dendro}.

\section{discussion}
\label{sec:discussion}
\subsection{Comparison of the \ammonia\ and \coeighteen}
\label{sec:comp_tracer}

Figure~\ref{fig:NH3_CO_integ_add_dendro} shows the integrated intensity of the \ammonia\ (1,1) main quadrupole hyperfine line superposed on the integrated intensity map of the FUGIN \coeighteen. The spatial distributions of the two emission lines appear to be consistent in most of our observed area, but in particular in the clumps 4 and 6 (IRAS 07077--1026 and IRAS 07081--1028) the distributions are different. The spatial range of the distributional discrepancy is about $\sim$1~pc.

\begin{figure}[t]
	\includegraphics[width=\linewidth]{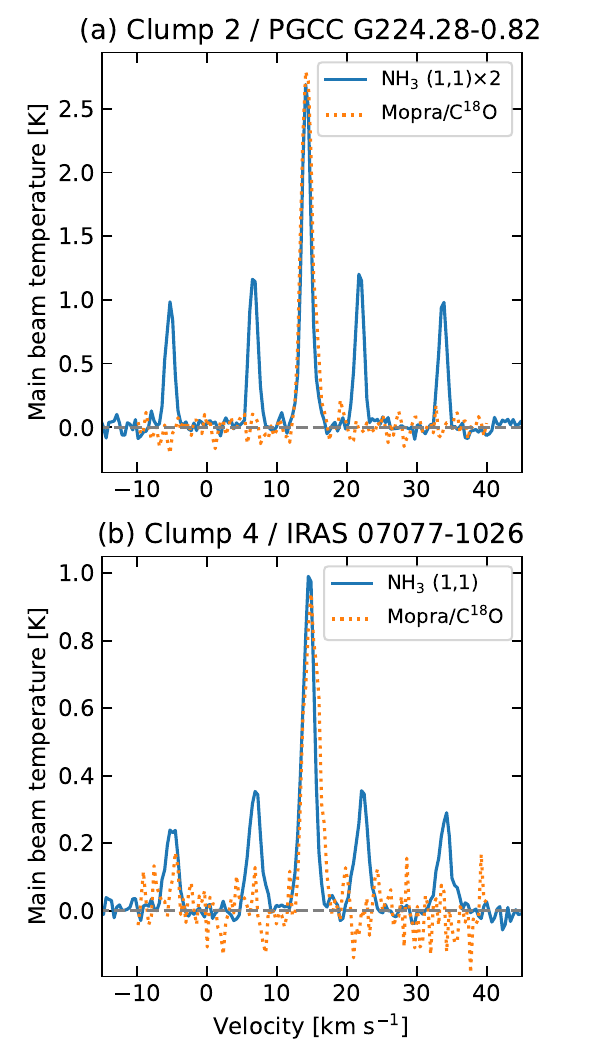}
    \caption{\ammonia\ (1,1) and Mopra \coeighteen\ ($J$=1--0) line profiles of the clumps 2 and 4. The spectra are located at the position (a) and (b) in figure \ref{fig:target_source}. For comparison, both molecular line data were smoothed to the spectral resolution of 0.38 \kms\, and Mopra \coeighteen\ ($J$=1--0) data was smoothed to a beam size of 75\arcsec. (a): Emission line profiles of the clump 2. For easy comparison of line widths, the \ammonia\ (1,1) intensity is scaled by a factor of 2. (b): Emission line profiles of the clump 4. (Color online)}
    \label{fig:compare_NH3_C18O_profile}
\end{figure}

If there is a difference for the gas density traced by the \ammonia\ (1,1) and \coeighteen\ ($J$=1--0) emission lines, this could explain the distributional discrepancy. However, we believe these two lines trace almost the same molecular gas because of their line profiles. Figure \ref{fig:compare_NH3_C18O_profile} shows the \ammonia\ (1,1) and Mopra \coeighteen\ ($J$=1--0) spectra. Two panels show the spectra at the clumps 2 and 4, corresponding to positions (a) and (b) in figure \ref{fig:target_source}. For easy comparison, the Mopra \coeighteen\ data was smoothed to be the same spatial and velocity resolutions of the \ammonia\ data. Although the width of the \coeighteen\ is slightly wider than that of the \ammonia\ (1,1) at either position, the differences are close to the velocity resolution. The two molecular emission lines are likely to trace almost identical layers in the molecular cloud.

\begin{figure}
    \includegraphics[width=\linewidth]{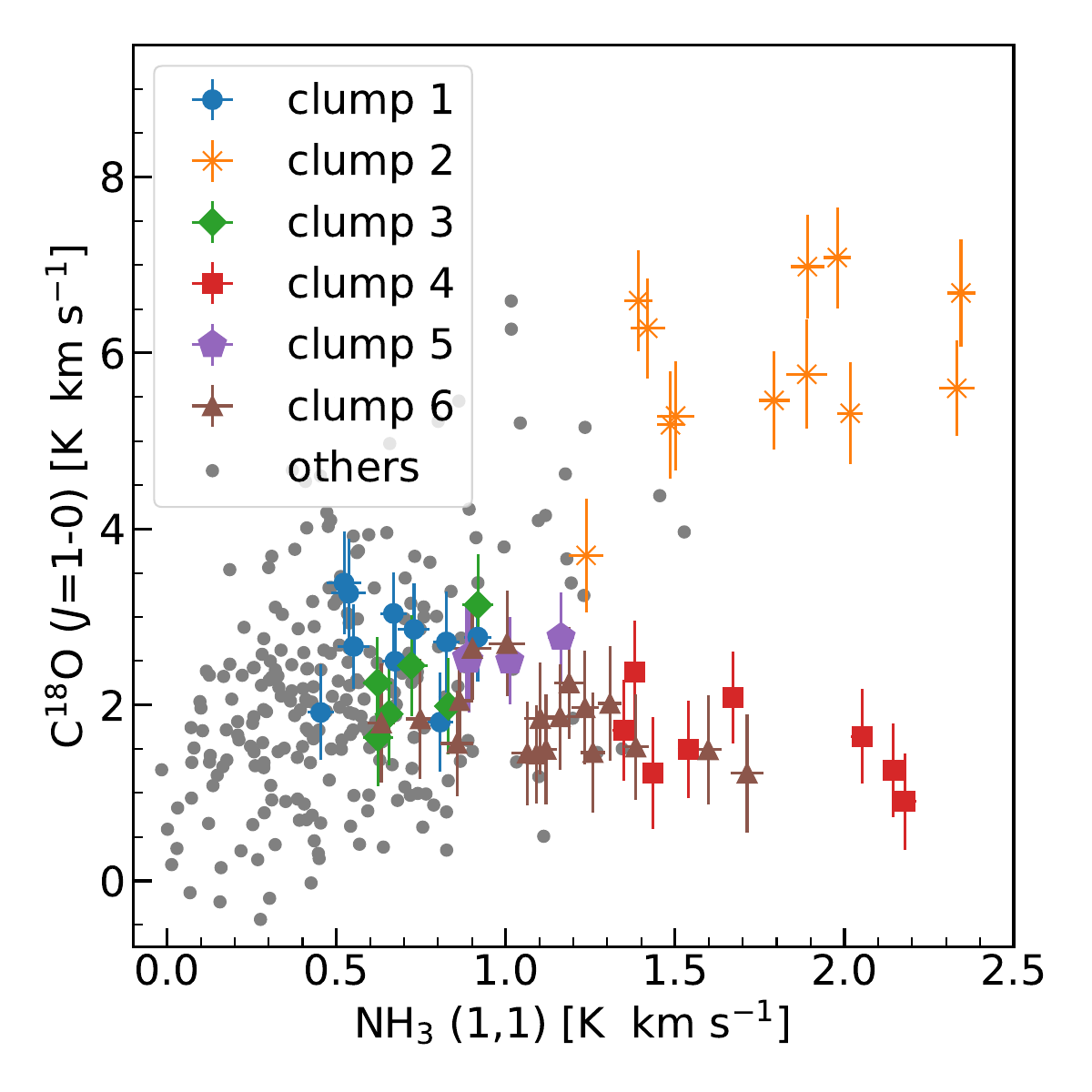}
    \caption{\ammonia\ (1,1) -- \coeighteen\ ($J$=1--0) integrated intensity correlation plot of the each clump and the other positions. (Reprinted with permission from \cite{Handa_2023}. Copyright The Author(s), 2023 Published by Cambridge University Press on behalf of International Astronomical Union.) (Color online)
    }
    \label{fig:NH3_CO_integ_correlation}
\end{figure}

To investigate the distributional discrepancy more quantitatively, we made an integrated intensity correlation plot (figure~\ref{fig:NH3_CO_integ_correlation}). In this plot, we used the pixels contained in the clumps identified by the dendrogram analysis. 
We can find two features in figure~\ref{fig:NH3_CO_integ_correlation}: a positive correlation in the clump 2, with a correlation coefficient of $r = 0.42$, while negative correlations are found in the clumps 4 and 6, with $r$ of $-$0.56 and $-$0.41, respectively.
Since the two lines are expected to trace identical layers in the cloud, our results suggest that the \coeighteen\ emission line intensity is suppressed compared to the \ammonia\ emission in the clumps 4 and 6.

The distributional discrepancy between the integrated intensity of the \coeighteen\ ($J$=1--0) and the $N$(H$_2$) or the integrated intensity of the \ammonia\ (1,1) could have several causes; including CO depletion, photodissociation where the molecules are destroyed by UV radiation, and high-density gas dissipation. These possibilities will be discussed in subsection~\ref{sec:posi_weak_C18O}.

\subsection{Potential Error of \coeighteen\ optical depth}
\label{sec:tau_uncertainty}

The optical depth has a strong influence on the accuracy of column density estimates.
In particular, when $\tau$ $>$ 1, deriving the reliable column density is difficult because molecular lines cannot trace inside the cloud.
Using the brightness temperatures of CO and \coeighteen\ lines with a 38\arcsec\ beam, we derived that the beam averaged optical depths of \coeighteen\ line are 0.15 - 0.44 and 0.05 - 0.15 for the clumps 2 and 4, respectively. It means the estimated column densities are reliable. However, the actual optical depths must be larger in the case of the beam filling factor is less than unity, as a result, the beam averaged column density may be underestimated. It is important to check that the mean optical depth of the emitting region is less than unity.

In the case of \ammonia\ (1,1) line, the estimated optical depth is free from the beam filling factor, because it is estimated from the intensity ratio of the quadrupole hyperfine lines at the same transition.
Therefore, we can estimate the beam filling factor for the \ammonia\ (1,1) line, $\phi_\mathrm{NH_3}$, assuming that $T_\mathrm{ex}$ = $T_\mathrm{rot}$.
The mean $\phi_\mathrm{NH_3}$ over the clump are about 0.15 and 0.09 for the clumps 2 and 4, respectively, for a 75\arcsec\ beam. It means the beam filling factor of Mopra \coeighteen\ ($J$=1--0) is 0.60 and 0.34 for the clumps 2 and 4, respectively, with a 38\arcsec\ beam, if the size of \coeighteen\ ($J$=1--0) emitting area is the same as that of \ammonia\ (1,1).

Using these values of the beam filling factor, the mean \coeighteen\ optical depths over the emitting region are 0.30 - 0.90 and 0.14 - 0.50 for the clumps 2 and 4, respectively. In any case, our estimation of the beam averaged column density of \coeighteen\ is reliable enough.

\subsection{Cause of weak \coeighteen\ emission in dense clump}
\label{sec:posi_weak_C18O}

%------------------------------------------------
In subsection~\ref{sec:comp_tracer}, we showed two regions where the weak \coeighteen\ emission line is detected despite high H$_2$ column density. This phenomenon in the clump 4 was also reported by \cite{Olmi_2016}. They described two possible causes for this phenomenon: CO depletion and differences in the temperature dependence of CO and dust emissions. It has also been slightly discussed by \cite{Sewi_o_2019}, together with the detection of the H$_2$O maser, the high density gas has been interpreted as a decrease in the clump 4. However, this phenomenon is hardly discussed in the previous studies and the reason for this discrepancy is not identified. In this subsection we will discuss the cause of this discrepancy, and the clump 4 will be treated as the main subject of discussion unless otherwise stated. 

\cite{Sewi_o_2019} suggested that the gas in the the clump 4 may be decreasing by dissipation. In general, if the gas density is sufficiently lower than the critical density, the molecular emission line cannot be detected. However, $\rm{N_{2}H^{+}}$ ($J$=1--0) emission line was detected in this region \citep{Tatematsu_2017}. The critical density of $\rm{N_{2}H^{+}}$ ($J$=1--0) is $4.1 \times 10^4$ \ccm\ at $T$ = 20 K \citep{Shirley_2015}, which is an order of magnitude higher than that of \coeighteen\ ($J$=1--0). It is therefore unlikely that the cause of the discrepancy is a decrease in the gas density. Furthermore, the luminosity-mass ratio, $L_\mathrm{bol}/\MO$, of three dust clumps associated with the region within a 1.$\arcmin$5 radius of the central position of the clump 4 was less than 1.0 \citep{Elia_2013}. Sources with $L_\mathrm{bol}/\MO$ $<$ 1 contain either low-mass protostars or early-stage intermediate-mass/high-mass protostars \citep{Molinari_2016}. Also, the H$_2$O maser does not reflect a well-defined evolutionary stage of the host YSOs \citep{Titmarsh_2016}. Therefore, the cause of the discrepancy is unlikely to be dense gas dispersion.

We considered that \coeighteen\ molecules are likely to be destroyed by  photodissociation \citep[e.g.,][]{Glassgold_1985}. Photodissociation is a phenomenon in which a molecule dissociates when it collides with a high-energy photon. Photodissociation is often detected near the boundary surface between \HII\ regions and molecular clouds. The two IRAS sources reported the candidate for an ultra-compact (UC) \HII\ region \citep{Bronfman_1996}. 
However, two IRAS sources are not listed in the catalogue of the \HII\ region using the Wide-Field Infrared Survey Explorer \citep[][]{Anderson_2014}. Furthermore, in and around two IRAS sources, there are no Red MSX Source objects, which are candidates for high-mass young stellar objects and Compact or UC \HII\ regions. Therefore, the clumps 2 and 4 are unlikely to contain a \HII\ region.
In addition, the dissociation energies of CO and \ammonia\ are 11.092 eV and 4.3 eV, respectively \citep{Dappen2002}. It is natural to assume that if a photodissociation region is formed, \ammonia\ molecules will be destroyed before \coeighteen\ molecules, making \ammonia\ lines undetectable. Furthermore, it is common that for dust clumps associated with \HII\ region to have $L_\mathrm{bol}/\MO \gtrsim 40$ \citep{Giannetti_2017}. Considering that dust clumps associated with the clumps 4 and 6 have $L_\mathrm{bol}/\MO$ $<$ 2, the cause of the discrepancy is unlikely to be the destruction of \coeighteen\ molecules by photodissociation.

We believe that CO depletion is the most likely cause of the discrepancy. Both the kinetic temperature and the dust temperature in and around the clumps 4 and 6 are lower than 20 K, which is temperature criterion for CO depletion. N$_2$H$^+$ which detected in the clump 4 immediately reacts with CO in the gas-phase to form HCO$^+$ \citep[e.g.,][]{Jorgensen_2004}. In other words, the detection of N$_2$H$^+$ implicitly indicates that CO depletion is occurring at least in the centre of the region. It is therefore suggested that the \coeighteen\ line cannot detect high density gas in the clump 4 due to the CO depletion.

Based on this idea, some questions remain as to why CO depletion occurs in the clump 4.
First, CO depletion is a temperature dependent phenomenon. It is somewhat unnatural that CO depletion occurs in the clump 4, even though both the dust temperature and the kinetic temperature are higher than the ambient temperature. However, in the clump 4, the dust temperature and the kinetic temperature are 14 K and 17 K, respectively, and CO depletion can occur under these conditions \citep[e.g.,][]{Pillai_2007,Fontani_2012,Feng_2020}. It is therefore not surprising that CO depletion occurs even at slightly higher temperatures than in the surrounding region. On the other hand, it is also questionable why regions other than clumps 4 and 6 are not affected by CO depletion. In subsection \ref{sec:depletion_condition}, a comparison of clump 2 and clump 4 provides a discussion of this question. Second, a SiO emission line is also detected in IRAS 07077--1026 \citep{Harju_1998}. This suggests that the outflow shock associated with a protostar destroys the icy mantle of dust grains. However, the outflow shock is highly directional and is unlikely to affect the entire clump. In fact, outflow and CO depletion have been found to occur simultaneously \citep[][]{Tobin_2013,Feng_2016,Feng_2016b}. Therefore, this idea is consistent with previous studies. 

It is unclear whether CO depletion occurs everywhere within the clump; \cite{Olmi_2023} reported that an ALMA observation of dust clump within the clump 4, named HG2825, has detected strong \coeighteen\ ($J$=2--1) emission.
\cite{Ge_2020} also proposed that there are multiple cores overlapping towards the line-of-sight in this region, forming a complex distribution of different molecular species. High-resolution CO observations over the entire clump are necessary to understand the inner structure.

The clump 6 shows similar features to the clump 4 on the integrated intensity correlation plot of the \ammonia\ (1,1) and \coeighteen\ ($J$=1--0) (figure~\ref{fig:NH3_CO_integ_correlation}).
However, due to the lack of N$_2$H$^+$ observations, it is unclear whether CO depletion occurs in the clump 6. Further observations are needed to verify this.

Our results suggest that the presence of pc-scale CO depletion can be detected by comparing the integrated intensity of \ammonia\ and \coeighteen, such as in low-mass star-forming regions \citep{Willacy_1998}. Since molecular clouds are concentrated in the Galactic plane and crowded towards the line of sight, it is difficult to compare the distributions of the dust continuum and \coeighteen\ line emission, and to accurately estimate the properties of depleted regions. Molecular emission lines can be separated using the velocity field information, and the intensities of the emission lines can be easily compared. In addition, the \ammonia\ line provides many physical parameters, such as gas temperature and column density (see subsection~\ref{sec:physic_params}). Comparing the intensity distributions of \ammonia\ and \coeighteen\ lines will allow us to identify many candidates for pc-scale CO depletion in the Galactic plane and to make statistical arguments about the physical conditions of the pc-scale CO depletion.

\subsection{\coeighteen\ depletion factor}
\label{sec:dep_factor}

It is a common practice to use the depletion factor, $f_\mathrm{D}$ \cite[e.g.,][]{Caselli_1999}, in order to evaluate the degree of CO depletion. The definition of $f_\mathrm{D}$ is as follows,
\begin{equation}
\label{eq:depletion_factor}
    f_\mathrm{D} = \frac{X^\mathrm{E}(\mathrm{C^{18}O})}{X^\mathrm{O}(\mathrm{C^{18}O})},
\end{equation}
where $X^\mathrm{E}$(\coeighteen) is the "expected" (i.e., canonical) relative abundance of \coeighteen\ molecules to molecular hydrogen and $X^\mathrm{O}$(\coeighteen) is the observed one. The \coeighteen\ abundance of the solar neighbourhood or the \coeighteen\ abundance predicted by the galactocentric distance are frequently used as the $X^\mathrm{E}$(\coeighteen).

The depletion factor obtained from equation~(\ref{eq:depletion_factor}) varies by a factor of several, depending on assumptions such as dust opacity. When discussing only a single molecular cloud, focusing on changes in the depletion factor within a cloud would reduce the impact of the underlying assumptions.
Therefore, we defined the "relative" depletion factor at the same molecular cloud, $f_\mathrm{D}^\mathrm{R}$:
\begin{equation}
\label{eq:relative_depletion_factor}
    f_\mathrm{D}^\mathrm{R} = \frac{X^\mathrm{M}(\mathrm{C^{18}O})}{X^\mathrm{O}(\mathrm{C^{18}O})},
\end{equation}
where $X^\mathrm{M}$(\coeighteen) is the maximum $X^\mathrm{O}$(\coeighteen) in the observed region. We calculated \coeighteen\ relative abundance, $X\rm{(C^{18}O)}$, and the $X\rm{(C^{18}O)}$ in the \ammonia\ observed area ranges from $4 \times 10^{-8}$ to $3.7 \times 10^{-7}$. To improve reliability, the value of $X^\mathrm{M}$(\coeighteen) is determined from observed pixels that represent more than 10 times the error of $N$(\coeighteen), and we adopt $2.5 \times 10^{-7}$. 

\begin{figure}
	\includegraphics[width=\linewidth, trim={0 0 0 0}, clip]{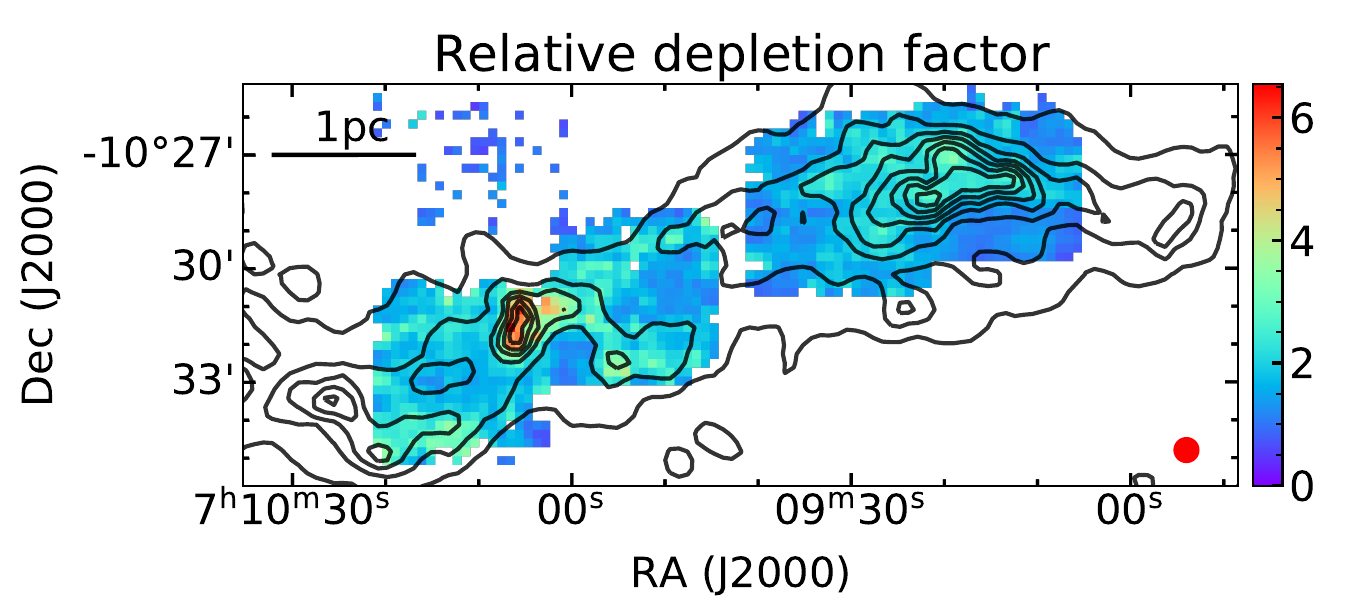}
    \caption{The spatial distribution of the relative depletion factor, $f_\mathrm{D}^\mathrm{R}$. The contours are the same as in figure~\ref{fig:Herschels}-(a). The red filled circle in the bottom right corner indicates the beam size of the Mopra data. (Color online)
    }
    \label{fig:depletion_fact}
\end{figure}

Figure~\ref{fig:depletion_fact} shows the spatial distribution of $f_\mathrm{D}^\mathrm{R}$ derived from the Mopra data. 
The $f_\mathrm{D}^\mathrm{R}$ is about 3.5 to 6.5 in and around the clump 4, and about 1.5 to 3.5 in and around the clump 2.
The freeze-out rate of $\mathrm{C^{18}O}$ is about 2 to 4 times higher in the clump 4 than in the clump 2.

We found the pc-scale CO depletion in the clump 4 (figure~\ref{fig:depletion_fact}). \cite{Feng_2020} have reported that 70 \micron\ dark clumps with pc-scale CO depletion have a depletion factor about 2 to 4 times larger than 70 \micron\ bright clumps without CO depletion at the same IRDCs. The trend of CO depletion observed in the clump 4 is similar to that of previous studies. While these 70 \micron\ dark clumps fulfil the empirical threshold for high-mass star formation \citep{Kauffmann_2010}, the clump 4 is a relatively low-mass clump and is below the empirical threshold. A relatively low-mass clump that cannot form high-mass stars was found to have the same trend of pc-scale CO depletion as high-mass clumps.

\subsection{The condition of weak CO depletion in the centre of PGCC G224.28--0.82}
\label{sec:depletion_condition}

From figure~\ref{fig:depletion_fact}, we can see that the CO depletion state is different within the observed region. The star-forming clumps located in the observed region can be regarded as having substantially the same external physical conditions and initial conditions for the dense gas formation. It would be important to study the differences in the CO depletion state; the differences in the CO depletion state of the clumps may affect the chemical diversity of the pre/proto-stellar core, since the possibility has been suggested that prestellar cores grow by mass accretion from the surroundings during the protostar formation process \citep{Takemura_2021}. In this subsection, we focus on the differences in physical properties between the clump 2 and clump 4.

To investigate the difference in CO depletion state, we compared the temperature and volume density of these clumps. In the clump 2, $T_{\mathrm{dust}}$ and $T_{\mathrm{kin}}$ are both approximately 12 K, and the mean gas density, calculated assuming the clump is a symmetric sphere, is estimated to be $4 \times 10^{4}$ \ccm. 
In the clump 4, $T_{\mathrm{dust}}$ $\simeq$ 14 K, $T_{\mathrm{kin}}$ $\simeq$ 16 K, and the mean gas density is estimated to be 3 $\times$ 10$^{4}$ \ccm.

These estimates fulfil the conditions for the occurrence of CO depletion. In other words, if CO depletion has occurred in the clump 4, it is natural that CO depletion should also occur in the clump 2. The clump 2 fulfils the empirical threshold for high-mass star formation, and intuitively, clump 2 is more likely to cause pc-scale CO depletion than clump 4. Since the gas and dust temperatures at the clump 4 are higher than at the clump 2, and the clumps 2 and 4 also have similar densities, this difference is not caused by temperature and density. In the clump 2, there is only weak CO depletion. Therefore the \coeighteen\ ($J$=1--0) emission can trace the dense gas. This implies that what determines the conditions under which \coeighteen\ ($J$=1--0) emission line can and cannot trace dense gas is unclear.

\begin{figure}
	\includegraphics[width=\linewidth]{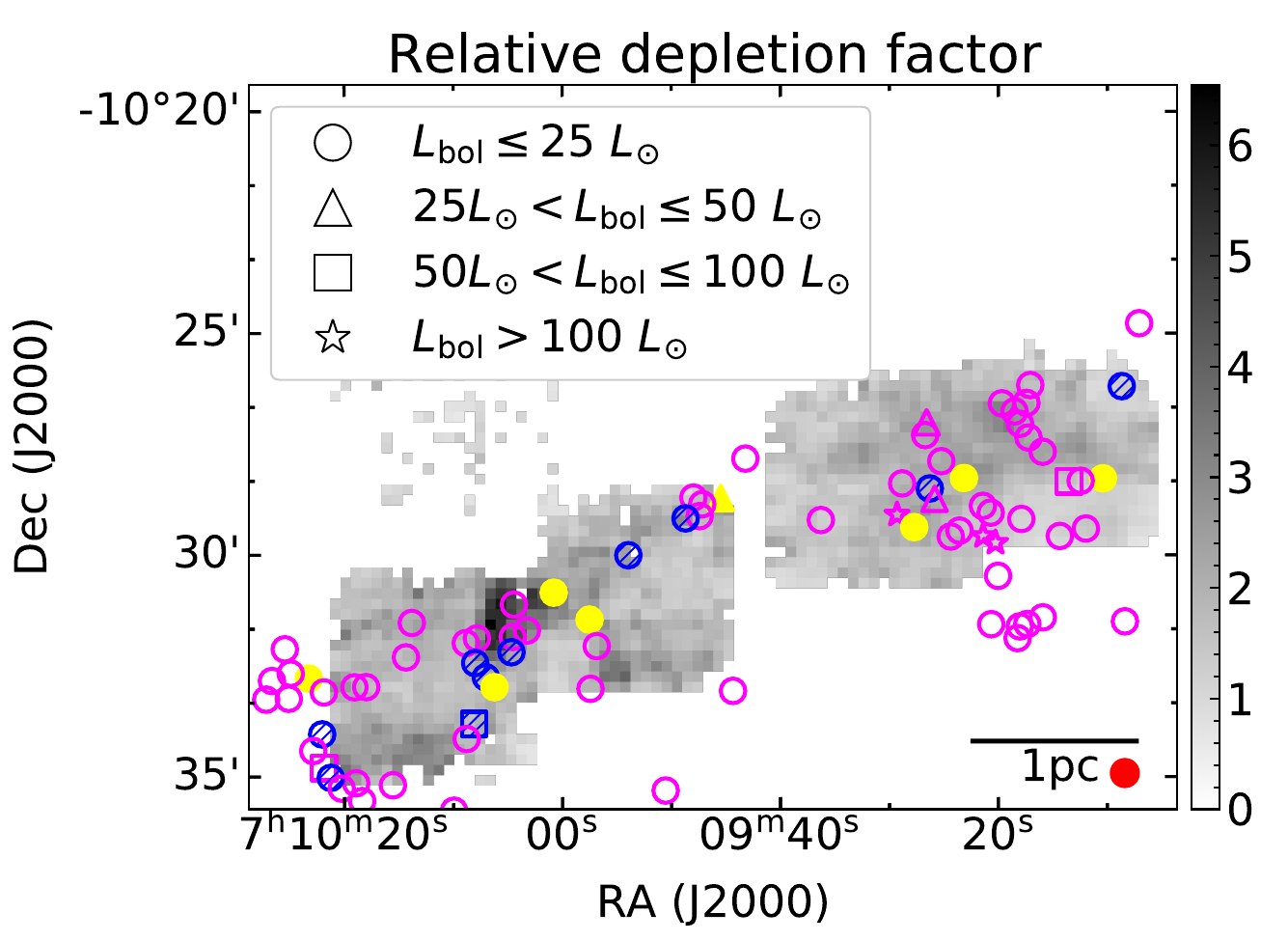}
    \caption{The locations of the protostar candidates (plots) superposed on the relative depletion factor map (colour image). Magenta open, blue hatched, and yellow filled markers indicate ``\textit{disk-only}'' sources, ``\textit{envelope and disk}'' sources, and ``\textit{envelope-only}'' sources, respectively. The red filled circle in the bottom right corner indicates the beam size of the Mopra data. (Color online)}

    \label{fig:relative_dep_fac_yso}
\end{figure}

This phenomenon is possibly caused by non-thermal desorption such as photodesorption, chemical sputtering. Molecules frozen onto dust grains may have been released into the gas phase by photodesorption at the clump 2. Figure~\ref{fig:relative_dep_fac_yso} shows the spatial distribution of $f_\mathrm{D}^\mathrm{R}$ and the locations of the protostar candidates reported by \cite{Sewi_o_2019}, classified by bolometric luminosity and evolutionary stages of YSO. As this figure shows, there are three protostars with bolometric luminosity above 100 $\LO$ in the vicinity of only the clump 2. 
These protostars have only a disk and with no envelope, and can affect the surrounding environment.
Furthermore, even protostars deeply embedded in molecular clouds may be able to influence the surrounding environment, if molecular clouds have clumpy structures as proposed by previous studies \cite[e.g.,][]{Spaans_1996,Kramer_2008,Shimajiri_2014}; as a result, molecules frozen onto dust grains are released into the gas phase by photodesorption.

However, photodesorption is well known to occur in the cold outer regions of proto-planetary disks \citep[e.g.,][]{Oberg_2015} as well as in the centre of the pre-stellar core \citep[e.g.,][]{Caselli_2012}, but it is not known in a cold clump at the pc scale. Therefore, our proposal needs further verification. In order to study the effects of photodesorption by radiation from protostars on the surrounding environment, observations of molecular species such as H$_2$O, whose abundance increases with photodesorption, are essential.

\section{Conclusions}
\label{sec:conclutions}
We carried out mapping observations towards the star-forming region, which is the vicinity of ($\alpha_\mathrm{J2000},\delta_\mathrm{J2000}$) = (07:09:20.30, $-$10:27:55) associated with CMa OB1 in the \ammonia\ (1,1), (2,2), (3,3), and H$_2$O maser emission lines using the Nobeyama 45 m radio telescope.
Our results are summarised as follows:
\begin{enumerate}
    \item Our observations revealed physical parameters of the observed region such as $T_\mathrm{kin}$ and $N$(\ammonia). The $T_\mathrm{kin}$ ranges from 10.2 K to 17.3 K, with IRAS 07077--1026 having a higher temperature by about 3 K higher than the other regions. The total column density of \ammonia\ does not vary significantly in most of the regions. The mean value is $N$(\ammonia) = $1.4 \times 10^{15}$ \scm.
    \item We carried out a dendrogram analysis using \ammonia\ (1,1) main quadrupole hyperfine line data and identified six clumps. The centre of PGCC G224.28--0.82 corresponds to the clump 2, IRAS 07077--1026 and IRAS 07081--1028 correspond to the clumps 4 and 6, respectively. The clump radius ranges from 0.19 pc to 0.39 pc and clump mass ranges from 41 $\MO$ to 290 $\MO$. Only the clump 2 fulfils the empirical threshold for high-mass star formation.
    \item By comparing the spatial distributions of the \ammonia\ (1,1) integrated intensity and the \coeighteen\ ($J$=1--0) integrated intensity, we found pc-scale intensity anti-correlations in the clump 4 and clump 6. Moreover, the comparison of these integrated intensity distributions with the distribution of $N$(H$_2$) obtained from the dust shows that the \ammonia\ molecular emission lines reflect the distribution of $N$(H$_2$) well, even at the pc scale. On the contrary, the intensity distribution of the \coeighteen\ emission lines does not correlate well with the distribution of $N$(H$_2$), and some high-density regions cannot be detected at the pc scale. By examining the dissipation of the high-density gas, photodissociation, and CO depletion, we suggest that the reason why \coeighteen\ ($J$=1--0) does not trace $N$(H$_2$) is due to CO depletion in the clump 4. A comparison of the intensity distributions of the \ammonia\ and \coeighteen\ lines will allow us to identify candidates for pc-scale CO depletion.
    \item We calculated the relative depletion factor, $f_\mathrm{D}^\mathrm{R}$, and compared the $f_\mathrm{D}^\mathrm{R}$ of high H$_2$ column density regions, the clump 2 and clump 4. There are differences in $f_\mathrm{D}^\mathrm{R}$ by a factor of about 2 to 4. This is comparable to the difference in depletion factor between the 70 \micron\ dark clump and the 70 \micron\ bright clump in the same molecular cloud reported by \cite{Feng_2020}.
    \item We discussed why there is weak CO depletion in the clump 2. The difference in temperature and gas density does not explain this phenomenon well and the conditional differences that \coeighteen\ ($J$=1--0) can and cannot trace dense gas are unclear. Non-thermal desorption such as photodesorption may play a key role in changing the CO depletion state.
\end{enumerate}

%%%%%%%%%%%%%%%%%%%%%%%%%%%%%%%%%%%%%%%

\begin{ack}
The 45 m radio telescope is operated by the Nobeyama Radio Observatory, a branch of the National Astronomical Observatory of Japan. This publication makes use of data from FUGIN, FOREST Unbiased Galactic plane Imaging survey with the Nobeyama 45 m telescope, a legacy project in the Nobeyama 45 m radio telescope.
Figure~\ref{fig:NH3_CO_integ_add_dendro} and Figure~\ref{fig:NH3_CO_integ_correlation} are reprinted with permissoin from Handa et al., Parsec scale CO depletion in KAGONMA 71, or a star-forming filament in CMa OB1, Proceedings of the International Astronomical Union, 17(S373), 31-34, 2023. (Copyright The Author(s), 2023 Published by Cambridge University Press on behalf of International Astronomical Union.)
This research made use of \textsc{aplpy}, an open-source plotting package for Python \citep{aplpy2012}, \textsc{astropy},\footnote{\url{http://www.astropy.org}} a community-developed core Python package for Astronomy \citep{astropy_2013, astropy_2018}, \textsc{matplotlib}, a Python package for visualization \citep{Hunter_2007}, \textsc{numpy}, a Python package for scientific computing \citep{harris_2020}, \textsc{pandas} a Python package for statistical analysis, data manipulations and processing \citep{pandas}, \textsc{scipy}, a Python package for fundamental algorithms for scientific computing \citep{Virtanen2020}, and Overleaf, a collaborative tool.
We would like to thank the Nobeyama Radio Observatory staff members (NRO) for their assistance and observation support. We also thank the students of Kagoshima University for their support in the observations.
\end{ack}

\appendix 
\section{Intensity differences of \co}
\label{sec:intensity_diff_12co}
\begin{figure}
	\includegraphics[width=\linewidth, trim={0 0 0 0}, clip]{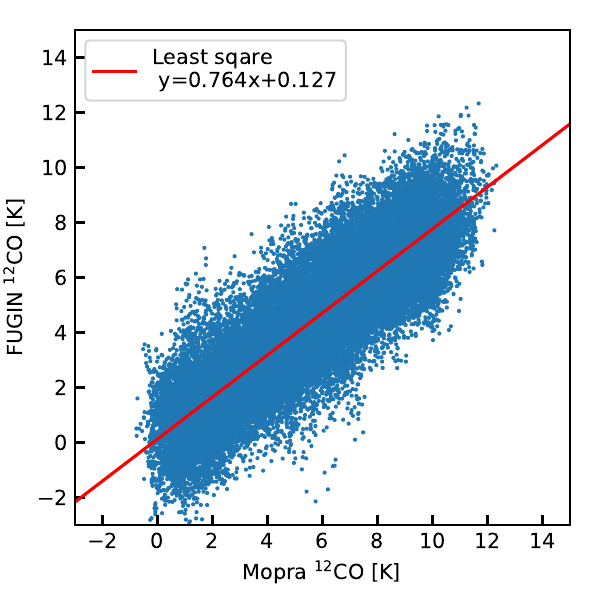}
    \caption{Intensity correlation between the Mopra \co\ and the FUGIN \co. The data from the 11 channels in the range from 11.7 \kms\ to 18.2 \kms\ are plotted. The red line shows the best-fit linear function determined by a least-squares method. (Color online)
    }
    \label{fig:intensity_diff_12co}
\end{figure}

Figure~\ref{fig:intensity_diff_12co} shows the intensity correlation between the Mopra \co\ and the FUGIN \co. The FUGIN \co\ data was smoothed to the beam size of the Mopra data and regridded to the grid size of the Mopra data. The Mopra \co\ data was smoothed to the velocity resolution of the FUGIN data and was interpolated onto the FUGIN data grid. The resulting rms noise of the FUGIN \co\ is 1.2 K and that of the Mopra \co\ is 0.30 K. We plot the data for 11 channels ranging from 11.7 \kms\ to 18.2 \kms. They fit well with a linear function of $T_\mathrm{mb}$(FUGIN \co) = 0.764 $\times$ $T_\mathrm{mb}$(Mopra \co) + 0.127; the red line overlaid on the data shows the best-fit linear function. The intensity of the Mopra \co\ data is systematically 1/0.764 $\simeq$ 1.3 times greater than the intensity of the FUGIN \co\ data. We obtained the result $T_\mathrm{mb}$(FUGIN \coeighteen) = 0.973 $\times$ $T_\mathrm{mb}$(Mopra \coeighteen) $-$ 0.038 from the similar analysis using the \coeighteen\ data. There is no significant difference in intensity between the FUGIN \coeighteen\ and the Mopra \coeighteen. It is not known whether the problem is in the FUGIN \co\ data or the Mopra \co\ data and what the cause might be. The following paragraph describes the effect on the physical parameters, in the case of the intensity of the Mopra \co\ data was 30\% higher than in reality.

\begin{figure}
	\includegraphics[width=\linewidth, trim={0 0 0 0}, clip]{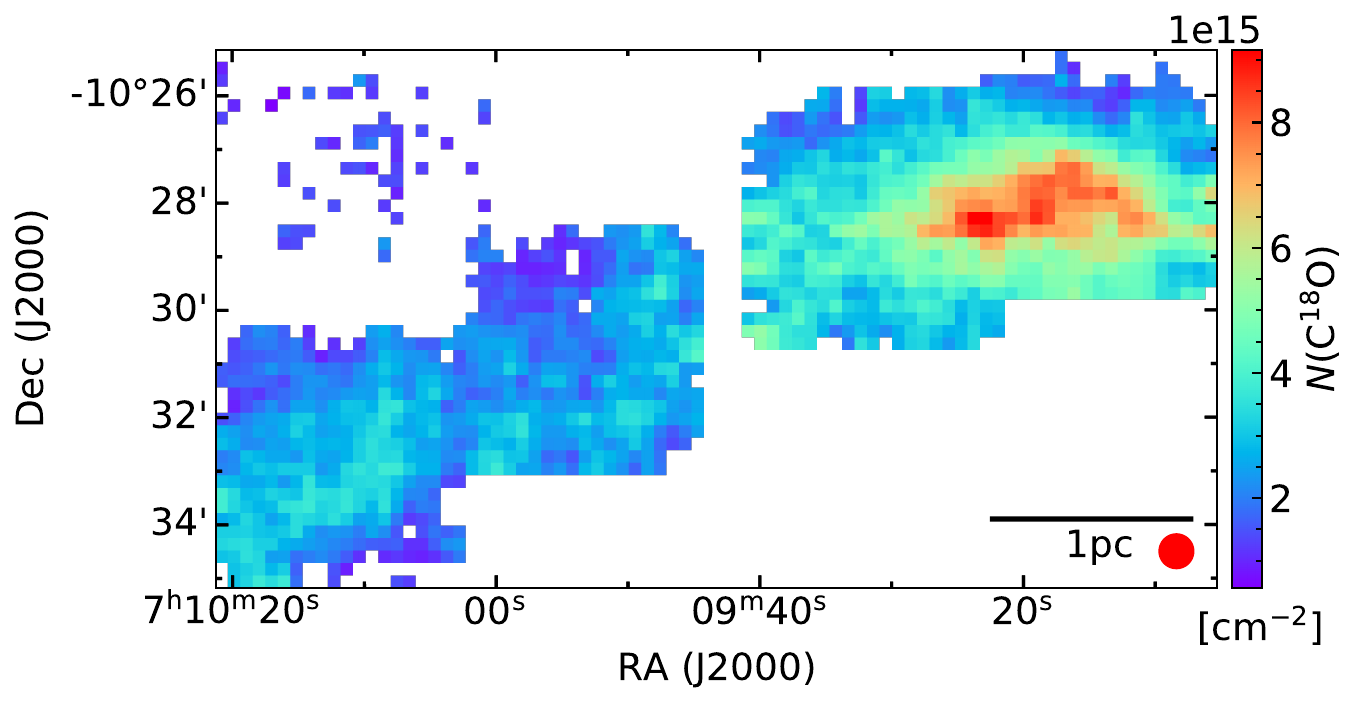}
    \caption{Same as figure~\ref{fig:C18O_column} but for the re-calculated \coeighteen\ column density map derived from the Mopra \co\ data with the intensity scaled by a factor of 0.764. (Color online)
    }
    \label{fig:re-calculated_c18o_column}
\end{figure}

\begin{figure}
	\includegraphics[width=\linewidth, trim={0 0 0 0}, clip]{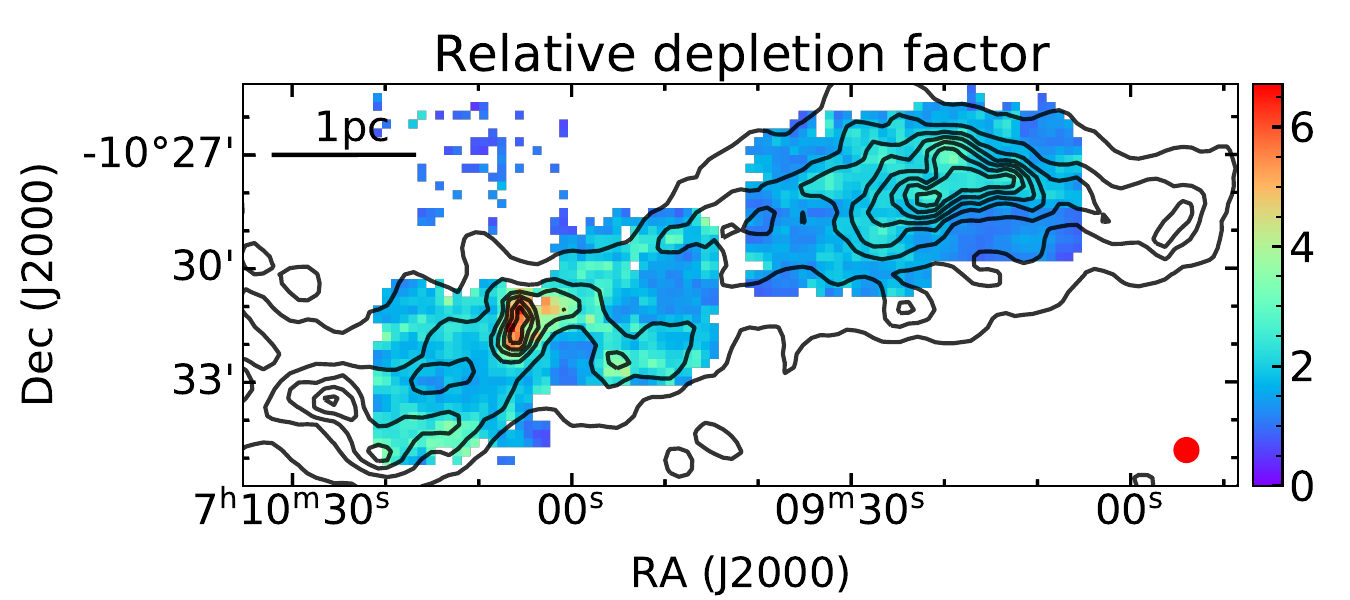}
    \caption{Same as figure~\ref{fig:depletion_fact} but for the re-calculated relative depletion factor derived from the re-calculated \coeighteen\ column density. (Color online)
    }
    \label{fig:re-calculated_dep_fac}
\end{figure}

We re-calculated $T_\mathrm{ex}$, $\tau_\mathrm{C^{18}O}$, $N$(\coeighteen), $X$(\coeighteen), and $f_\mathrm{D}^\mathrm{R}$. The intensity of the Mopra \co\ was scaled by a factor of 0.764 and the following physical parameters were derived without changing other conditions. The $T_\mathrm{ex}$ changes by a factor of 0.81 to 0.84. The $\tau_\mathrm{C^{18}O}$ changes by a factor of 1.3 to 1.4. The $N$(\coeighteen) and the $X$(\coeighteen) change by a factor of 0.91 to 1.0. The $f_\mathrm{D}^\mathrm{R}$ changes by a factor of 0.94 to 1.0. Figure~\ref{fig:re-calculated_c18o_column} shows the re-calculated $N$(\coeighteen) and figure~\ref{fig:re-calculated_dep_fac} shows the re-calculated $f_\mathrm{D}^\mathrm{R}$. Both the values and the spatial distributions are not significantly different from those before the re-calculation.

\section{Correlations between \ammonia\ (1,1) and Dust emission}
\label{sec:correlations_NH3_Dust}
\begin{figure}
	\includegraphics[width=\linewidth]{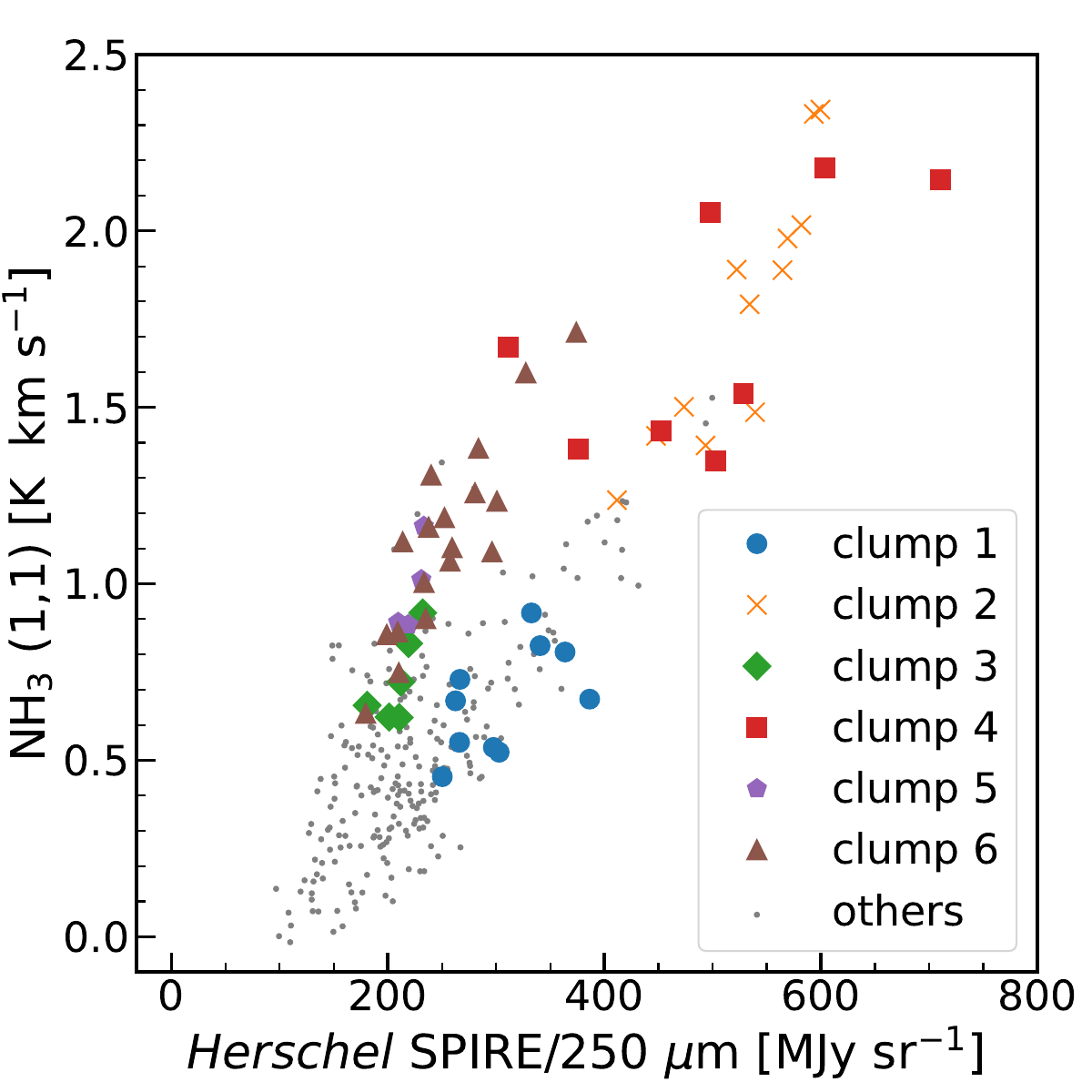}
    \caption{
    The correlation diagram between the \ammonia\ integrated intensity and the dust emission intensity at 250 \micron. The dust emission data was smoothed to a beam size of 75\arcsec. All points are colour coded for each clump and non-clump. (Color online)
    }
    \label{fig:correlations_NH3_Dust}
\end{figure}
Figure~\ref{fig:correlations_NH3_Dust} is the correlation plot between the \ammonia\ (1,1) integrated intensity and the dust emission intensity at 250 \micron. The correlation coefficient between these two data across all pixels is 0.83. There is a clear positive correlation and the same trend regardless of the strength of the two emissions.

\section{H$_2$O maser spectrum}
\label{sec:maser_spectrum}
\begin{figure*}
	\includegraphics[width=\linewidth, trim={110 0 110 10}, clip]{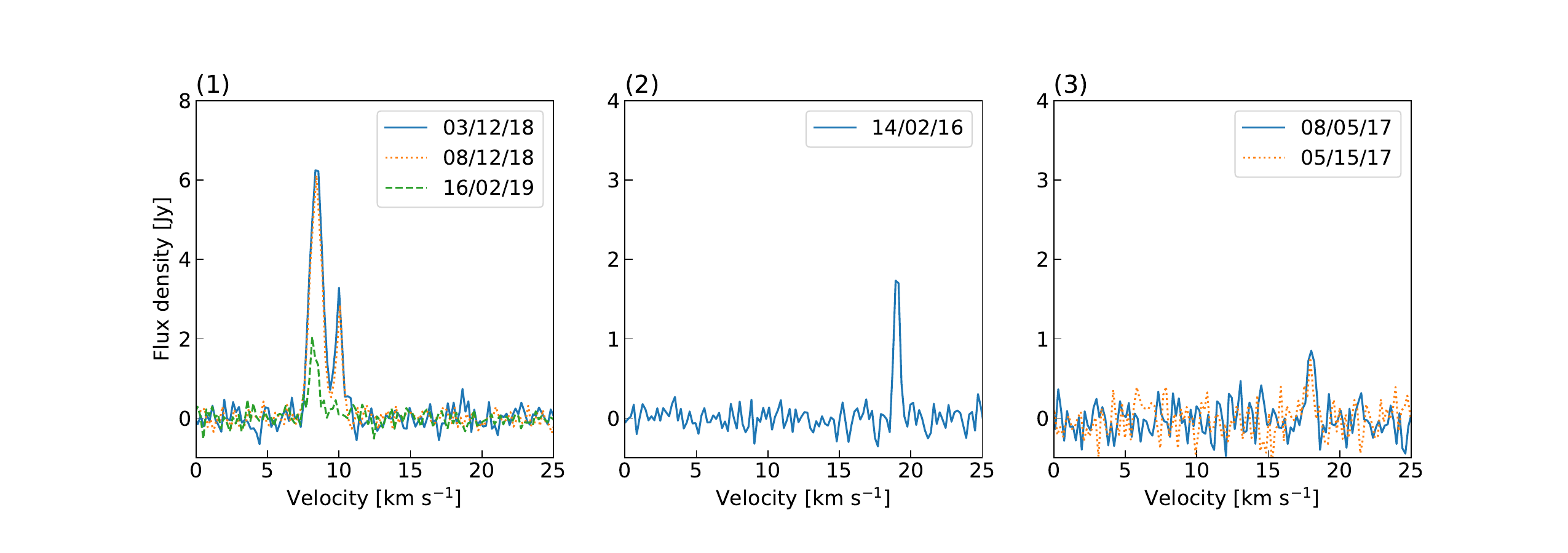}
    \caption{The H$_2$O maser emission from our observations. The detected dates and positions are listed in table~\ref{tab:water_maser}. (Color online)
    }
    \label{fig:maser_spectra}
\end{figure*}

Figure~\ref{fig:maser_spectra} shows the spectra of the H$_2$O masers listed in table~\ref{tab:water_maser}.

%%%
% See the manual for the detail.
%%%


\begin{thebibliography}{}
\bibitem[Anderson et~al.(2014)]{Anderson_2014}
{Anderson}, L.~D.,  {Bania}, T.~M.,  {Balser}, D.~S.,  {Cunningham}, V., {Wenger},
  T.~V., {Johnstone}, B.~M., \& {Armentrout}, W.~P. 2014, \apjs, 212, 1

\bibitem[Astropy Collaboration et~al.(2013)]{astropy_2013}
{Astropy Collaboration}, {et~al.} 2013, \aap, 558, A33

\bibitem[Astropy Collaboration et~al.(2018)]{astropy_2018}
{Astropy Collaboration}, {et~al.} 2018, \aj, 156, 123

\bibitem[{{Bergin} \& {Tafalla}(2007)}]{Bergin_2007}
{Bergin}, E.~A., \& {Tafalla}, M. 2007, \araa, 45, 339

\bibitem[Brand et~al.(1994)]{Brand_1994}
{Brand}, J., {et~al.} 1994, \aaps, 103, 541

\bibitem[Bronfman et~al.(1996)]{Bronfman_1996}
{Bronfman}, L., {Nyman}, L.~A., \& {May}, J. 1996, \aaps, 115, 81

\bibitem[Caselli et~al.(1999)]{Caselli_1999}
{Caselli}, P., {Walmsley}, C.~M., {Tafalla}, M., {Dore}, L., \& {Myers}, P.~C.
  1999, \apjl, 523, L165

\bibitem[Caselli et~al.(2012)]{Caselli_2012}
{Caselli}, P., {et~al.} 2012, \apjl, 759, L37

\bibitem[Clari{\'a}(1974)]{Claria_1974}{Clari{\'a}}, J.~J. 1974, \aap, 37, 229

\bibitem[Crapsi et~al.(2007)]{Crapsi_2007}
{Crapsi}, A., {Caselli}, P., {Walmsley}, M.~C., \& {Tafalla}, M. 2007, \aap, 470, 221

\bibitem[Dame et~al.(2001)]{Dame_2001}
{Dame}, T.~M., {Hartmann}, D., \& {Thaddeus}, P. 2001, \apj, 547, 792

\bibitem[Danby et~al.(1988)]{Danby_1988}
{Danby}, G., {Flower}, D.~R., {Valiron}, P., {Schilke}, P., \& {Walmsley},
  C.~M. 1988, \mnras, 235

\bibitem[{D{\"a}ppen(2002)}]{Dappen2002}
D{\"a}ppen, W. 2002, in Allen's Astrophysical Quantities, ed. A.~N. Cox (New York:
  Springer), 27--51

\bibitem[Dhabal et~al.(2019)]{Dhabal_2019}
{Dhabal}, A., {Mundy}, L.~G., {Chen}, C.~-y., {Teuben}, P., \& {Storm}, S. 2019, \apj, 876, 108

\bibitem[Elia et~al.(2013)]{Elia_2013}
{Elia}, D., {et~al.} 2013, \apj, 772, 45

\bibitem[Feng et~al.(2016a)]{Feng_2016}
{Feng}, S., {Beuther}, H., {Zhang}, Q., {Henning}, T., {Linz}, H., {Ragan}, S., \& {Smith}, R. 2016a, \aap, 592,
  A21

\bibitem[Feng et~al.(2016b)]{Feng_2016b}
{Feng}, S., {Beuther}, H., {Zhang}, Q., {Liu}, H.~B., {Zhang}, Z., {Wang}, K., \& {Qiu}, K.
  2016b, \apj, 828, 100

\bibitem[Feng et~al.(2020)]{Feng_2020}
{Feng}, S., {et~al.} 2020, \apj, 901, 145

\bibitem[Fischer et~al.(2016)]{Fischer_2016}
{Fischer}, W.~J., {Padgett}, D.~L., {Stapelfeldt}, K.~L., \& {Sewi{\l}o}, M.
  2016, \apj, 827, 96

\bibitem[Fontani et~al.(2012)]{Fontani_2012}
{Fontani}, F., {Giannetti}, A., {Beltr{\'a}n}, M.~T., {Dodson}, R., {Rioja}, M.,
  {Brand}, J., {Caselli}, P., \& {Cesaroni}, R. 2012, \mnras, 423, 2342

\bibitem[Frerking et~al.(1982)]{Frerking_1982}
{Frerking}, M.~A., {Langer}, W.~D., \& {Wilson}, R.~W. 1982, \apj, 262, 590

\bibitem[Ge et~al.(2020)]{Ge_2020}{Ge}, J.~X., {et~al.} 2020, ApJ, 891, 36

\bibitem[Giannetti et~al.(2017)]{Giannetti_2017}
{Giannetti}, A., {Leurini}, S., {Wyrowski}, F., {Urquhart}, J., {Csengeri}, T., {Menten}, K.~M., {K{\"o}nig}, C., \& {G{\"u}sten}, R. 2017, \aap, 603, A33

\bibitem[Glassgold et~al.(1985)]{Glassgold_1985}
{Glassgold}, A.~E., {Huggins}, P.~J., \& {Langer}, W.~D. 1985, \apj, 290, 615

\bibitem[Gong et~al.(2018)]{Gong_2018}
{Gong}, Y., {Li}, G.~X., {Mao}, R.~Q., {Henkel}, C., {Menten}, K.~M., {Fang}, M., {Wang}, M., \& {Sun}, J.~X. 2018, \aap, 620, A62

\bibitem[Griffin et~al.(2010)]{Griffin_2010}{Griffin}, M.~J., {et~al.} 2010, \aap, 518, L3

\bibitem[Handa {et~al.}(2023)]{Handa_2023}
{Handa}, T., {et~al.} 2023, IAU Symposium, 373, 31

\bibitem[Harju et~al.(1998)]{Harju_1998}
{Harju}, J., {Lehtinen}, K., {Booth}, R.~S., \& {Zinchenko}, I. 1998, \aaps, 132, 211

\bibitem[Harris et~al.(2020)]{harris_2020}{Harris}, C.~R., {et~al.} 2020, Nature, 585, 357

\bibitem[Hernandez et~al.(2011)]{Hernandez_2011}
{Hernandez}, A.~K., {Tan}, J.~C., {Caselli}, P., {Butler}, M.~J., {Jim{\'e}nez-Serra}, I., {Fontani}, F., \& {Barnes}, P. 2011, \apj, 738, 11

\bibitem[Hildebrand(1983)]{Hildebrand_1983}{Hildebrand}, R.~H. 1983, QJRAS, 24, 267

\bibitem[Ho et~al.(1977)]{Ho_1977}
{Ho}, P.~T.~P., {Martin}, R.~N., {Myers}, P.~C., \& {Barrett}, A.~H. 1977,
  \apjl, 215, L29

\bibitem[{{Ho} \& {Townes}(1983)}]{Ho_1983}{Ho}, P.~T.~P., \& {Townes}, C.~H. 1983, \araa, 21, 239

\bibitem[{Hunter(2007)}]{Hunter_2007}Hunter, J.~D. 2007, Computing in Science \& Engineering, 9, 90

\bibitem[Jim{\'e}nez-Serra et~al.(2014)]{Jimenez-Serra_2014}
{Jim{\'e}nez-Serra}, I., {Caselli}, P., {Fontani}, F., {Tan}, J.~C., {Henshaw}, J.~D., {Kainulainen}, J., \& {Hernandez}, A.~K. 2014, \mnras, 439, 1996

\bibitem[J{\o}rgensen et~al.(2004)]{Jorgensen_2004}
{J{\o}rgensen}, J.~K., {Sch{\"o}ier}, F.~L., \& {van Dishoeck}, E.~F. 2004, \aap, 416, 603

\bibitem[{{Kaltcheva} \& {Hilditch}(2000)}]{Kaltcheva_2000}{Kaltcheva}, N.~T., \& {Hilditch}, R.~W. 2000, \mnras, 312, 753

\bibitem[Kauffmann et~al.(2008)]{Kauffmann_2008}
{Kauffmann}, J., {Bertoldi}, F., {Bourke}, T.~L., {Evans}, N.~J., I., \& {Lee}, C.~W. 2008, \aap, 487, 993

\bibitem[{{Kauffmann} \& {Pillai}(2010)}]{Kauffmann_2010}{Kauffmann}, J., \& {Pillai}, T. 2010, \apjl, 723, L7

\bibitem[Kim et~al.(2004)]{Kim_2004}{Kim}, B.~G., {Kawamura}, A., \& {Fukui}, Y. 2004, \pasj, 56, 313

\bibitem[Kohno et~al.(2022)]{Kohno_2022}{Kohno}, M., {et~al.} 2022, \pasj, 74, 545
\bibitem[Kohno et~al.(2023)]{Kohno_2023}{Kohno}, M., {et~al.} 2023, \pasj, 75, 397

\bibitem[K{\"o}nyves et~al.(2015)]{Konyves_2015}{K{\"o}nyves}, V., {et~al.} 2015, \aap, 584, A91

\bibitem[Kramer et~al.(1999)]{Kramer_1999}
{Kramer}, C., {Alves}, J., {Lada}, C.~J., {Lada}, E.~A., {Sievers}, A., {Ungerechts}, H., \& {Walmsley}, C.~M. 1999, \aap, 342, 257

\bibitem[Kramer et~al.(2008)]{Kramer_2008}{Kramer}, C., {et~al.} 2008, \aap, 477, 547

\bibitem[Kuno et~al.(2011)]{kuno_2011}
Kuno, N., {et~al.} 2011, in 2011 XXXth URSI General
  Assembly and Scientific Symposium, 1--4

\bibitem[Ladd et~al.(2005)]{Ladd_2005}
{Ladd}, N., {Purcell}, C., {Wong}, T., \& {Robertson}, S. 2005, \pasa, 22, 62

\bibitem[Lewis et~al.(2021)]{Lewis_2021}
{Lewis}, J.~A., {Lada}, C.~J., {Bieging}, J., {Kazarians}, A., {Alves}, J., \& {Lombardi}, M. 2021, \apj, 908, 76

\bibitem[Lin et~al.(2021)]{Lin_2021}
{Lin}, Z., {Sun}, Y., {Xu}, Y., {Yang}, J., \& {Li}, Y. 2021, \apjs, 252, 20

\bibitem[Lombardi et~al.(2014)]{Lombardi_2014}
{Lombardi}, M., {Bouy}, H., {Alves}, J., \& {Lada}, C.~J. 2014, \aap, 566, A45

\bibitem[{{Mangum} \& {Shirley}(2015)}]{Mangum_2015}
{Mangum}, J.~G., \& {Shirley}, Y.~L. 2015, \pasp, 127, 266

\bibitem[Mangum et~al.(1992)]{Mangum_1992}
{Mangum}, J.~G., {Wootten}, A., \& {Mundy}, L.~G. 1992, \apj, 388, 467

\bibitem[Molinari et~al.(2016)]{Molinari_2016}
{Molinari}, S., {Merello}, M., {Elia}, D., {Cesaroni}, R., {Testi}, L., \& {Robitaille}, T. 2016, \apjl, 826, L8

\bibitem[Molinari et~al.(2010)]{Molinari_2010}{Molinari}, S., {et~al.} 2010, \aap, 518, L100

\bibitem[Murase et~al.(2022)]{Murase_2021}
{Murase}, T., {Handa}, T., {Hirata}, Y., {Omodaka}, T., {Nakano}, M., {Sunada}, K., {Shimajiri}, Y., \& {Nishi}, J. 2022, \mnras, 510, 1106

\bibitem[{{Myers} \& {Benson}(1983)}]{Myers_1983}{Myers}, P.~C., \& {Benson}, P.~J. 1983, \apj, 266, 309

\bibitem[Nagahama et~al.(1998)]{Nagahama_1998}{Nagahama}, T., {Mizuno}, A., {Ogawa}, H., \& {Fukui}, Y. 1998, \aj, 116, 336

\bibitem[{\"O}berg et~al.(2015)]{Oberg_2015}
{{\"O}berg}, K.~I., {Furuya}, K., {Loomis}, R., {Aikawa}, Y., {Andrews}, S.~M., {Qi}, C., {van Dishoeck}, E.~F., \& {Wilner}, D.~J. 2015, \apj, 810, 112

\bibitem[Olmi et~al.(2023)]{Olmi_2023}{Olmi}, L., {Brand}, J., \& {Elia}, D. 2023, \mnras, 518, 1917

\bibitem[Olmi et~al.(2016)]{Olmi_2016}
{Olmi}, L., {Cunningham}, M., {Elia}, D., \& {Jones}, P. 2016, \aap, 594, A58

\bibitem[Pety et~al.(2017)]{Pety_2017}{Pety}, J., {et~al.} 2017, \aap, 599, A98

\bibitem[Pickett et~al.(1998)]{Pickett_1998}
{Pickett}, H.~M., {Poynter}, R.~L., {Cohen}, E.~A., {Delitsky}, M.~L., {Pearson}, J.~C., \& {M{\"u}ller}, H.~S.~P. 1998, \jqsrt, 60, 883

\bibitem[Pilbratt et~al.(2010)]{Pilbratt_2010}{Pilbratt}, G.~L., {et~al.} 2010, \aap, 518, L1

\bibitem[Pillai et~al.(2007)]{Pillai_2007}
{Pillai}, T., {Wyrowski}, F., {Hatchell}, J., {Gibb}, A.~G., \& {Thompson}, M.~A. 2007, \aap, 467, 207

\bibitem[Pineda et~al.(2008)]{Pineda_2008}{Pineda}, J.~E., {Caselli}, P., \& {Goodman}, A.~A. 2008, \apj, 679, 481

\bibitem[Planck Collaboration et~al.(2011)]{Planck_2011_XXV}{Planck Collaboration}, {et~al.} 2011, \aap, 536, A25

\bibitem[Planck Collaboration et~al.(2016a)]{Planck_XXVIII_2016}
{Planck Collaboration}, {et~al.} 2016a, \aap, 594, A28

\bibitem[Planck Collaboration et~al.(2016b)]{Planck_48_2016}
{Planck Collaboration}, {et~al.} 2016b, A\&A, 596, A109

\bibitem[Poglitsch et~al.(2010)]{Poglitsch_2010}{Poglitsch}, A., {et~al.} 2010, \aap, 518, L2

\bibitem[Ripple et~al.(2013)]{Ripple_2013}
{Ripple}, F., {Heyer}, M.~H., {Gutermuth}, R., {Snell}, R.~L., \& {Brunt}, C.~M. 2013, \mnras, 431, 1296
\bibitem[{{Robitaille} \& {Bressert}(2012)}]{aplpy2012}
{Robitaille}, T., \& {Bressert}, E. 2012, {APLpy: Astronomical Plotting Library
  in Python}, {ascl}:{1208.017}

\bibitem[Roccatagliata et~al.(2015)]{Roccatagliata_2015}
{Roccatagliata}, V., {et~al.} 2015, \aap, 584, A119

\bibitem[Rosolowsky et~al.(2008)]{Rosolowsky_2008}
{Rosolowsky}, E.~W., {Pineda}, J.~E., {Kauffmann}, J., \& {Goodman}, A.~A. 2008, \apj, 679, 1338

\bibitem[Rydbeck et~al.(1977)]{Rydbeck_1977}
{Rydbeck}, O.~E.~H., {Sume}, A., {Hjalmarson}, A., {Ellder}, J., {Ronnang}, B.~O., \& {Kollberg}, E. 1977, \apjl, 215, L35

\bibitem[Sabatini et~al.(2019)]{Sabatini_2019}
{Sabatini}, G., {Giannetti}, A., {Bovino}, S., {Brand}, J., {Leurini}, S., {Schisano}, E., {Pillai}, T., \& {Menten}, K.~M. 2019, \mnras, 490, 4489

\bibitem[Sabatini et~al.(2022)]{Sabatini_2022}{Sabatini}, G., {et~al.} 2022, \apj, 936, 80,

\bibitem[Schisano et~al.(2014)]{Schisano_2014}{Schisano}, E., {et~al.} 2014, \apj, 791, 27

\bibitem[Sewi{\l}o et~al.(2019)]{Sewi_o_2019}{Sewi{\l}o}, M., {et~al.} 2019, ApJS, 240, 26

\bibitem[Shimajiri et~al.(2014)]{Shimajiri_2014}{Shimajiri}, Y., {et~al.} 2014, \aap, 564, A68

\bibitem[Shirley(2015)]{Shirley_2015}{Shirley}, Y.~L. 2015, \pasp, 127, 299

\bibitem[Sipil{\"a} et~al.(2019)]{Sipila_2019}
{Sipil{\"a}}, O., {Caselli}, P., {Redaelli}, E., {Juvela}, M., \& {Bizzocchi},
  L. 2019, \mnras, 487, 1269

\bibitem[Sokolov et~al.(2017)]{sokolov_2017}{Sokolov}, V., {et~al.} 2017, \aap, 606, A133

\bibitem[{{Spaans}(1996)}]{Spaans_1996}{Spaans}, M. 1996, \aap, 307, 271

\bibitem[Sunada et~al.(2007)]{Sunada_2007}
{Sunada}, K., {Nakazato}, T., {Ikeda}, N., {Hongo}, S., {Kitamura}, Y., \& {Yang}, J. 2007, \pasj, 59, 1185
\bibitem[Swift et~al.(2005)]{Swift_2005}
{Swift}, J.~J., {Welch}, W.~J., \& {Di Francesco}, J. 2005, \apj, 620, 823

\bibitem[Tafalla et~al.(2002)]{Tafalla2002}
{Tafalla}, M., {Myers}, P.~C., {Caselli}, P., {Walmsley}, C.~M., \& {Comito}, C. 2002, \apj, 569, 815

\bibitem[Takemura et~al.(2021)]{Takemura_2021}{Takemura}, H., {et~al.} 2021, \apjl, 910, L6

\bibitem[Tatematsu et~al.(2017)]{Tatematsu_2017}{Tatematsu}, K., {et~al.} 2017, \apjs, 228, 12

\bibitem[Titmarsh et~al.(2016)]{Titmarsh_2016}
{Titmarsh}, A.~M., {Ellingsen}, S.~P., {Breen}, S.~L., {Caswell}, J.~L., \&
  {Voronkov}, M.~A. 2016, \mnras, 459, 157

\bibitem[Tobin et~al.(2013)]{Tobin_2013}{Tobin}, J.~J., {et~al.} 2013, \apj, 765, 18

\bibitem[{{Togi} \& {Smith}(2016)}]{Togi_2016}
{Togi}, A., \& {Smith}, J.~D.~T. 2016, \apj, 830, 18

\bibitem[{{Ulich} \& {Haas}(1976)}]{Ulich_1976}
{Ulich}, B.~L., \& {Haas}, R.~W. 1976, \apjs, 30, 247

\bibitem[Umemoto et~al.(2017)]{Umemoto_2017}{Umemoto}, T., {et~al.} 2017, \pasj, 69, 78

\bibitem[Urquhart et~al.(2011)]{Urquhart_2011}{Urquhart}, J.~S., {et~al.} 2011, \mnras, 418, 1689

\bibitem[Virtanen et~al.(2020)]{Virtanen2020}Virtanen, P., {et~al.} 2020, Nature Methods, 17, 261

\bibitem[Wang et~al.(2019)]{Wang_2019}
{Wang}, C., {Yang}, J., {Su}, Y., {Du}, F., {Ma}, Y., \& {Zhang}, S. 2019, \apjs, 243, 25

\bibitem[Wang et~al.(2020)]{wang_2020}
{Wang}, S., {Ren}, Z., {Li}, D., {Kauffmann}, J., {Zhang}, Q., \& {Shi}, H. 2020, \mnras, 499, 4432

\bibitem[{{W}es {M}c{K}inney(2010)}]{pandas}
{W}es {M}c{K}inney. 2010, in {P}roceedings of the 9th {P}ython in {S}cience
  {C}onference, ed. {S}t\'efan van~der {W}alt \& {J}arrod {M}illman, 56 -- 61

\bibitem[Willacy et~al.(1998)]{Willacy_1998}
{Willacy}, K., {Langer}, W.~D., \& {Velusamy}, T. 1998, \apjl, 507, L171

\bibitem[Zhou et~al.(2020)]{Zhou_2020}
{Zhou}, D.-d., {Wu}, G., {Esimbek}, J., {Henkel}, C., {Zhou}, J.-j., {Li}, D.-l., {Ji}, W.-g., \& {Zheng}, X.-w. 2020, \aap, 640, A114
\end{thebibliography}
\end{document}